\documentclass{CSML}
\pdfoutput=1

\usepackage{lastpage}
\newcommand{\dOi}{13(4:14)2017}
\lmcsheading{}{1--\pageref{LastPage}}{}{}%
{Oct.~28, 2016}{Nov.~22, 2017}{}

\usepackage[english]{babel}
\usepackage[hidelinks]{hyperref}
\hypersetup{hidelinks}
\usepackage{listings}
\usepackage{amsfonts,amsmath,amssymb}
\usepackage{xspace}
\usepackage{dsfont}
\usepackage{algorithm}

\theoremstyle{plain}

\lstdefinelanguage{coq}
{
  morekeywords ={Definition, Lemma, Theorem, forall, exists, Inductive,
    CoInductive, Type, Class, Hypothesis},
  sensitive=true,
  morecomment =[s]{(*}{*)},
  escapeinside={(@}{@)},
  emph={Prop}, emphstyle=\bf
}
\newcommand{\ls}[1]{\lstinline{#1}}

\newcommand{\nde}{{process\/}}

\newcommand{\neighbors}[1]{\mathcal{N}_{#1}}

\newcommand{\IsShort}{\mbox{\textit{IsShort\/}}}
\newcommand{\IsTall}{\mbox{\textit{IsTall\/}}}

\newcommand{\kDominator}{\mbox{\textit{kDominator\/}}}

\newcommand{\Alpha}{\mbox{\textit{Alpha\/}}}
\newcommand{\TallChildren}{\mbox{\textit{TallChildren\/}}}
\newcommand{\ShortChildren}{\mbox{\textit{ShortChildren\/}}}
\newcommand{\MinIDMinATall}{\mbox{\textit{MinCMinATall\/}}}
\newcommand{\MinATall}{\mbox{\textit{MinATall\/}}}
\newcommand{\MaxAShort}{\mbox{\textit{MaxAShort\/}}}
\newcommand{\Algo}{\mathcal A}
\newcommand{\Coq}{Coq\xspace}

\newcommand{\algocl}[1]{\mathcal{C}({#1})} 

\newcommand{\parent}[1]{\mathtt{Par}({#1})}

\newcommand{\clrdom}[0]{\ensuremath{Dom}}
\newcommand{\cache}[1]{}

 \newcommand{\dsP}{{\mathds{P}}}
 \newcommand{\dsS}{{\mathds{S}}}

\renewcommand{\r}{\ensuremath{\textit{r}}}

\newcommand{\algoCLR}[1]{\ensuremath{\algocl{#1}}}
\newcommand{\parCl}{\ensuremath{parC}}
\newcommand{\head}{\ensuremath{headC}}
\newcommand{\ParCLR}{\ensuremath{ParC}}
\newcommand{\HeadCLR}{\ensuremath{HeadC}}
\newcommand{\Ids}{\ensuremath{Ids}}
\newcommand{\id}{\mathtt{Id}}

\newcommand{\Z}{\ensuremath{\mathbb{Z}}}

\newcommand{\ltM}{\ensuremath{\prec}}
\newcommand{\eqM}{\ensuremath{\approx}}
\newcommand{\neqM}{\ensuremath{\not\approx}}

\makeatletter
\def\subsubsection{\@startsection{subsubsection}{3}%
  \z@{.5\linespacing\@plus.7\linespacing}{-.5em}%
  {\normalfont\bfseries}}

\def\paragraph{\@startsection{paragraph}{4}%
  \z@\z@{-\fontdimen2\font}%
  {\normalfont\bfseries}}
\makeatother

\begin{document}

\title[A Framework for Certified Self-Stabilization]{A Framework for
  Certified Self-Stabilization$^\star$}

\author[Altisen et al.]{Karine Altisen \and Pierre Corbineau \and St\'ephane Devismes}
\address{Univ. Grenoble Alpes, CNRS, Grenoble INP$^\dag$\footnote{$^\dag$ Institute of
Engineering Univ. Grenoble Alpes}, VERIMAG, 38000 Grenoble, France
}
\email{Firstname.Lastname@univ-grenoble-alpes.fr}



\titlecomment{$^\star$ This work is an extended version of
  \cite{APD16} and has been partially supported by ANR projects
  \textsc{Estate} (ANR-16-CE25-0009) and \textsc{Descartes}
  (ANR-16-CE40-0023).}

\begin{abstract}
  We propose a general framework to build certified proofs of
  distributed self-stabilizing algorithms with the proof assistant
  \Coq.
  We first define in \Coq the locally shared memory model with
  composite atomicity, the most commonly used model in the
  self-stabilizing area.
  We then validate our framework by certifying a non trivial part of
  an existing silent self-stabilizing algorithm which builds a
  $k$-clustering of the network. We also certify a quantitative
  property related to the output of this algorithm. Precisely, we show
  that the computed $k$-clustering contains at most $\lfloor
  \frac{n-1}{k+1} \rfloor + 1$ clusterheads, where $n$ is the number
  of nodes in the network.
   To obtain these results, we also developed a library which contains
   general tools related to potential functions and cardinality of
   sets.
\end{abstract}

\maketitle

\section{Introduction}
\label{sec:introduction}

In 1974, Dijkstra introduced the notion of {\em self-stabilizing\/}
algorithm~\cite{Dijkstra74} as any distributed algorithm that resumes
correct behavior within finite time, regardless of the initial
configuration of the system.  A self-stabilizing algorithm can
withstand \textit{any} finite number of transient faults.  Indeed,
after transient faults hit the system and place it in some arbitrary
configuration --- where, for example, the values of some variables 
have been arbitrarily modified --- a self-stabilizing
algorithm is guaranteed to resume correct behavior without external
({\em e.g.}, human) intervention within finite time.  Thus,
self-stabilization makes no hypothesis on the nature or extent of
transient faults that could hit the system, and recovers from the
effects of those faults in a unified manner.

For more than 40 years, a vast literature on self-stabilizing
algorithms has been developed. Self-stabilizing solutions have been
proposed for many kinds of classical distributed problems, {\em e.g.},
token circulation~\cite{HuangC93}, spanning tree
construction~\cite{CYH91}, clustering~\cite{CaronDDL10},
routing~\cite{Dolev1997122}, propagation of information with
feedback~\cite{BuiDPV99}, clock synchronization~\cite{CouvreurFG92},
{\em etc}. Moreover, self-stabilizing algorithms have been designed to
handle various environments, {\em e.g.}, wired
networks~\cite{HuangC93,CYH91,CaronDDL10,Dolev1997122,BuiDPV99,CouvreurFG92},
WSNs \cite{BenOthman2013199,Telematik_SSS_2013_Neighborhood},
peer-to-peer
systems~\cite{DBLP:journals/ppl/CaronDPT10,Caron20131533}, {\em etc}.

Progress in self-stabilization has led to consider more and more
adversarial environments. As an illustrative example, the three first
algorithms proposed by Dijkstra in 1974~\cite{Dijkstra74} were
designed for oriented ring topologies and assuming sequential
executions only, while nowadays most self-stabilizing algorithms
are designed for fully asynchronous arbitrary connected networks, {\em
  e.g.}, \cite{HuangC93,CaronDDL10,DLDHR12c}.

Consequently, the design of self-stabilizing algorithms becomes more
and more intricate, and accordingly, the proofs of their respective
correctness and complexity are now often tricky to establish. However,
proofs in distributed computing, in particular in
self-stabilization, are commonly written by hand, based on informal
reasoning. This potentially leads to errors when arguments are not
perfectly clear, as explained by Lamport in its position
paper~\cite{Lamport2012}. So, in the current context, such methods are
clearly pushed to their limits, since the question on confidence in
proofs naturally arises. 

This justifies the use of a {\em proof assistant}, a tool which allows
to develop certified proofs interactively and check them mechanically.
  In this paper, we use \Coq~\cite{coq}, recipient of the ACM 2013
  Software system Award. \Coq has been successfully employed for
  various tasks such as mathematical developments as involved in the
  Feit-Thompson theorem~\cite{gonthier13}, formalization of the
  correctness of a C compiler~\cite{leroy09jar,mccarthy67am},
  certified numerical libraries~\cite{perin+SAS2013}, and verification
  of cryptographic protocols~\cite{almeida12ccs,Corbineau+CPP2011}.

\subsection{Contribution}

We propose a general framework to build certified proofs of
self-stabilizing algorithms for wired networks with the tool \Coq.
We first define in \Coq the {\em locally shared memory model with
  composite atomicity}, introduced by Dijkstra~\cite{Dijkstra74}. This
model is the most commonly used in the self-stabilizing area. Our
modeling is versatile, {\em e.g.}, it supports any class of network
topologies (including arbitrary ones), the diversity of anonymity
levels (from fully anonymous to fully identified), and various levels
of asynchrony ({\em e.g.}, sequential, synchronous, fully
asynchronous).

We show how to use and validate our framework by certifying a
non trivial part of an existing silent self-stabilizing algorithm
proposed in \cite{DLDHR12c} which builds a $k$-clustering of the
network. Starting from an arbitrary configuration, a silent algorithm
converges within finite time to a configuration from which all
communication variables are constant. This class of self-stabilizing
algorithms is important, as self-stabilizing algorithms building
distributed data structures (such as spanning tree or clustering)
often achieve the silent property, and these silent self-stabilizing
data structures are widely used as basic building blocks for more
complex self-stabilizing solutions, {\em
  e.g.},~\cite{DLDHR12c,DLDHR13}.

Using a usual proof scheme, the certified proof consists of two
main parts, one dealing with termination and the other with partial
correctness.

For the termination part, we developed tools on potential functions
and termination at a fine-grained level.  Precisely, we define a
potential function as a multiset containing a local potential per
node. We then exploit two criteria that are sufficient to meet the
conditions for using the Dershowitz-Manna well-founded ordering on
multisets \cite{DershowitzM79}. These two criteria, and the
  associated proof scheme, are versatile enough to be applied to prove
  the termination many other (silent) algorithms, whether using our
  framework, or by hand.
We also provide tools to build termination proofs of algorithms
consisting of prioritized sets of actions. These tools use a 
lexicographical order on multisets of local potentials.
Notice that the termination proof we propose for the case study
assumes a distributed unfair daemon, the most general scheduling
assumption of the model. By contrast, the proof given in
\cite{DLDHR12c} uses a stronger scheduling hypothesis, namely, a
distributed weakly fair daemon.

The partial correctness part consists of showing that (1) a
$k$-clustering is defined in the network whenever the algorithm has
terminated, and (2) a quantitative property related to this
$k$-clustering; namely the computed $k$-clustering contains at most
$\lfloor \frac{n-1}{k+1} \rfloor + 1$ clusterheads, where $n$ is the
number of nodes in the network. To obtain this latter result, we
provide a library dealing with cardinality of sets in general and
properties on cardinals of finite sets {\em w.r.t.} basic set
operations, {\em i.e.}, Cartesian product, disjoint union and subsets.

This work is an extended version of \cite{APD16}; it represents about
17,560 lines of code (as computed by \ls{coqwc}: 5k lines of
specifications, 10k lines of proofs) written in \Coq\ 8.6 compiled
with OCaml 4.04.1.
%




\subsection{Related Work}  

  Many formal approaches have been used in the context of distributed
  computing.  There exist tools to validate a given distributed
  algorithm, such as the tools embedded with TLA+ (a model checker and
  a proof assistant) \cite{tla}.  Constructive approaches aim at
  synthesizing algorithms based on a given specification and a fixed
  topology; many of them are now based on SMT-solvers, see
  \cite{synth-cav16, synth-forte16}. Note that model-checking as well
  as synthesis are fully automated, but require to fix the topology and
  sometime the scheduling (\textit{e.g.}, synchronous
  execution). Moreover, these techniques usually succeed with small
  topologies, due to computation limits.  For example, model checking
  has been also successfully used to prove impossibility results
  applying on small-scale distributed systems~\cite{DLPRT12}.  In contrast, a
  proof assistant may validate a given algorithm for arbitrary-sized
  topologies, but is only semi-automated and requires heavy
  development for each algorithm.
We now focus on works related to certification of distributed
algorithms, most of them using \Coq.

Several works have shown that proof assistants (in particular \Coq)
are well-suited to certify the correction of algorithms as well as
impossibility results in various kinds of distributed
systems. Certification of non fault-tolerant (consequently non
self-stabilizing) distributed algorithms in \Coq is addressed
in~\cite{SCSS09,tase12,CourtieuRTU16}.  In \cite{CRTU15}, an
impossibility proof for the gathering problem is certified. Notice
that \cite{CRTU15,CourtieuRTU16} consider {\em mobile} distributed
systems. Precisely, these works are dedicated to swarms of robots that
are endowed with motion actuators and visibility sensors and deployed
in in the Euclidean plane. These robots are weak, {\em i.e.}, they are
anonymous, uniform, unable to explicitly communicate, and oblivious
(they have no persistent memory).

Certification in the context of fault-tolerant, yet non
self-stabilizing, distributed computing is addressed
in~\cite{isabelle,bouzid13sss}.  K\"ufner {\em et al.}~\cite{isabelle}
propose to certify (using the proof assistant {\em Isabelle})
fault-tolerant distributed algorithms. Their framework deals with
masking fault-tolerance whereas self-stabilization is non-masking by
essence. Moreover the network topology is restricted to fully
connected graphs. Bouzid {\em et al.}~\cite{bouzid13sss} certify
impossibility results for swarms of robots that are subjected to
Byzantine faults using a model based on the one described in
\cite{CRTU15}.


To the best of our knowledge, only three works deal with certification
of self-stabilizing algorithms~\cite{Courtieu02,deng09tase,PVS99}. 
First, \cite{deng09tase} proposes to certify in \Coq
self-stabilizing population protocols. Population protocols are used
as a theoretical model for a collection (or population) of tiny mobile
agents that interact with one another to carry out a computation. The
movement pattern of the agents is unpredictable, and communication is
implicit between close agents (there is no notion of communication
network).
%
A formal correctness proof of Dijkstra's seminal self-stabilizing
algorithm~\cite{Dijkstra74} is conducted with the PVS proof
assistant~\cite{PVS99}, where only sequential executions are
considered. In \cite{Courtieu02}, Courtieu proposes a setting for
reasoning on self-stabilization in \Coq. He restricts his study to
very simple self-stabilizing algorithms ({\em e.g.}, the 4-states
algorithm of Ghosh~\cite{G93}) working on networks of restrictive
topologies (lines and rings).  
%


\subsection{Roadmap} 

The rest of the paper is organized into two parts. The first one, from
Section~\ref{sec:model} to Section~\ref{sec:proof-counting-part},
describes the general framework. The case study is given in
Sections~\ref{sec:k-clust-algor}-\ref{sec:corr-algocl}.  Section
\ref{sec:ccl} is dedicated to concluding remarks and perspectives.

In the next section, we describe how we define the locally shared
memory model with composite atomicity in \Coq. In
Section~\ref{sec:silent_self_stab}, we express the definitions of
self-stabilization and silence in \Coq, moreover we certify a
sufficient condition to show that an algorithm is silent and
self-stabilizing.  We present tools for proving termination in
Section~\ref{sec:proof-convergence-part-general} and tools for proving
quantitative properties in Section~\ref{sec:proof-counting-part}.
In Section~\ref{sec:k-clust-algor}, we present an algorithm called
$\algocl{k}$ (the case study), its specification, and the assumptions
under which it will be proven. In Section~\ref{sec:term-algocl}, we
present the termination proof of
$\algocl{k}$. Section~\ref{sec:corr-algocl} deals with the partial
correctness of $\algocl{k}$.

Along this paper, we present our work together with few pieces of \Coq
code that we simplify in order to make them readable. In particular,
we intend to use notations as defined in the model or in the
algorithm.  The \Coq definitions, lemmas, theorems, and documentation
related to this paper are available as an online browsing at
\url{http://www-verimag.imag.fr/~altisen/PADEC/}. All source codes are
also available at this URL. We encourage the reader to visit this web
page for a deeper understanding of our work.



\section{Locally Shared Memory Model with Composite
  Atomicity}\label{sec:model}

In this section, we explain how we model the {\em locally shared
  memory model with composite atomicity} in \Coq. This model has been
introduced by Dijkstra~\cite{Dijkstra74}, and since then it is the
most commonly used in the self-stabilizing area.

\subsection{Distributed Systems} 

We define a \emph{distributed system} as a finite set of
interconnected nodes. Each node has its own private memory and
runs its own code. It can also interact with other nodes in the
network \textit{via} interconnections.
The model in \Coq reflects this by defining two independent
types:
\begin{itemize}
\item A \ls{Network} is equipped with a type \ls{Node}, representing
  nodes of the network. A \ls{Network} defines functions and
  properties that depict its topology, {\em i.e.}, interconnections
  between nodes. Those interconnections are specified using the type
  \ls{Channel}.

\item The \ls{Algorithm} of a node \ls{p} is equipped with a type
  \ls{State}, which describes the memory state of \ls{p}. Its main
  function, \ls{run}, specifies how \ls{p} executes and interacts with
  other nodes through channels (type \ls{Channel}).
\end{itemize}

\subsection{Network and Topology}
\label{sec:model:network}

Nodes in a distributed system can directly communicate with a subset
of other nodes. As commonly done in the literature, we view the
communication {\em network} as a simple directed graph $G = (V,E)$,
where $V$ is set of vertices representing nodes and $E\subseteq
V\times V$ is a set of edges representing direct communication between
distinct nodes. We write $n$ to denote the numbers of nodes: $n =
|V|$.

Two distinct nodes $p$ and $q$ are said to be {\em neighbors} if
$(p,q)\in E$. From a computational point of view, $p$ uses a distinct
channel $c_{p,q}$ to communicate with each of its neighbors $q$: it
does not have direct access to $q$. In the type \ls{Network}, the
topology is defined using this narrow point of view, \textit{i.e.},
interconnections (edges of the graph) are represented using channels
only.
In particular, the neighborhood of $p$ is encoded with the set
$\neighbors{p}$ which contains all channels $c_{p,q}$ outgoing from
$p$. The sets $\neighbors{p}$, for all $p$, are modeled in \Coq as
lists, using the function %
\lstinline{(peers: Node $\to$ list Channel)}.
The function %
\lstinline{(peer: Node $\to$ Channel $\to$ option Node)} returns the
destination neighbor for a given channel name, \textit{i.e.},
\lstinline{(peer $p$ $c_{p,q}$)} returns \lstinline{(Some $q$)}, or
$\bot$\footnote{Option type is used for partial functions which, by
  convention, return \ls{(Some _)} when defined, and \ls{None}
  otherwise. \ls{None} is denoted by $\bot$ in this paper.} if the name is
unused. We also define the shortcut ternary relation \ls{(is_channel p
  c p')} as \ls{(peer p c)} equals \ls{(Some p')} where \ls{p} and
\ls{p'} are nodes and \ls{c} a channel.

Communications can be made bidirectional, assuming a property called
\ls{sym_net}, which states that for all nodes $p_1$ and
$p_2$, the network defines a channel from $p_1$ to $p_2$ if and
only if it also defines a channel from $p_2$ to $p_1$.
In case of bidirectional links $(p,q)$ and $(q, p)$ in $E$, $p$ can
access its channel name at $q$ using the function
\lstinline{($\rho_p$: Channel $\to$ Channel)}. Thus, we have the
following identities: $\rho_p(c_{p,q})$ equals $c_{q,p} \in
\neighbors{q}$ and $\rho_q(c_{q,p})$ equals $c_{p,q} \in
\neighbors{p}$. In our \Coq model, the role of $\rho_p$ is assigned to
the function %
\lstinline{(reply_to: Node $\to$ Channel $\to$ Channel)}.

As a last requirement, we suppose that, since the number of nodes in the
network is finite, we have a list, called \ls{all_nodes}, containing
all the nodes.  In particular, this assumption makes the
\emph{emptiness test} decidable: this test states that for any
function \lstinline{(f: Node $\to$ option A)} (with \ls{A}, some
type), one can compute whether \ls{f} always returns $\bot$ for any
parameter.  This test is used in the framework to detect termination
of the algorithm.

As a means of checking actual usability of the \ls{Network} type
definition, we have defined a function that can build any finite
\ls{Network} from a description of its topology given by a list of
lists of neighbors.

\subsection{Computational Model}

In the \emph{locally shared memory model with composite atomicity},
nodes communicate with their neighbors using finite sets of locally
shared registers, called {\em variables}. A node can read its own
variables and those of its neighbors, but can only write to its own
variables.

\subsubsection{Distributed Algorithm}

Each node operates according to its local {\em program}.  A
\emph{distributed algorithm} $\Algo$ is defined as a collection of $n$
\emph{programs}, each operating on a single node.
The \emph{state} of a node in ${\Algo}$ is defined by the values of
its local variables and is represented using an abstract immutable
\Coq datatype \ls{State}. Such a datatype is usually implemented as a
record containing the values of the algorithm variables.
A node $p$ can access the states of its neighbors using the
corresponding channels: we call this the \emph{local configuration} of
$p$, and model it as a function typed \ls{(Local_Env :=
  Channel} $\to$ \ls{option State)} which returns the current state of a
neighbor, given the name of the corresponding channel (or $\bot$ for
an invalid name).

The program of each node $p$ in $\Algo$ consists of a finite set of
guarded actions:
$$
\langle guard \rangle\ \hookrightarrow \ \langle statement \rangle$$ %
The \emph{guard} is a Boolean expression involving variables of $p$
and its neighbors. The \emph{statement} updates some variables of $p$.
An action can be executed only if its guard evaluates to $true$; in
this case, the action is said to be \emph{enabled}.  A node is said to
be \emph{enabled} if at least one of its actions is enabled.  The
local program at node $p$ is modeled by a function \ls{run}
of type %
%
\lstinline{(State $\to$ }%
\lstinline{list Channel $\to$ }%
\lstinline{Local_Env $\to$ }
\lstinline{(Channel $\to$ Channel) $\to$ }%
\lstinline{option State)}.

This function accesses the local topology and states around $p$. It
takes as first argument the current state of $p$. The two other
arguments are $\neighbors{p}$ and $\rho_p$. These arguments allow the
function to access the local states of $p$'s neighbors.  The returned
value is the next state of node $p$ if $p$ is enabled, $\bot$
otherwise.
Note that \ls{run} provides a functional view of the
algorithm: it includes the whole set of possible actions, but returns
a single result; this model is thus restricted to \emph{deterministic
  algorithms}.\footnote{Finite non-determinism could be handled by having
  \ls{run} output \ls{(list} \ls{State)} instead of \ls{(option}
    \ls{State)}.}

\subsubsection{Semantics}

A \emph{configuration} \ls{g} of the system is defined as an instance
of the states of all nodes in the system, \textit{i.e.}, a function
typed %
\lstinline{(Env := Node $\to$ State)}. For a given node \ls{p} and a
configuration \ls{(g: Env)}, the term \ls{(g p)} represents the state
of \ls{p} in configuration \ls{g}. Thanks to this encoding, we easily
obtain the local configuration (type \ls{Local_Env}) of node \ls{p} by
composing \ls{g} and \ls{peer} as a function %
\ls{(local_env g p) :=} %
\ls{(fun (c: Channel)} \ls{=> option_map g} \ls{(peer p c))}, where
\ls{option_map g} \ls{(peer p c)} returns \ls{(g p')} when
\ls{(peer p c)} returns \ls{Some p'}, and $\bot$ otherwise. Hence, the
execution of the algorithm on node \ls{p} in the current configuration
\ls{g} is obtained by:
\lstinline{(run (g p) $\neighbors{\tt p}$ (local_env g p) $\rho_{\tt p}$)}; 
it returns either $\bot$ if the node is disabled or \ls{(Some s)}
where \ls{(s: State)} is the next state of \ls{p}.
We define \ls{(enabled_b g p)} as the Boolean value (type \ls{bool})
\ls{true} if node \ls{p} is enabled in configuration
\ls{g} and \ls{false} otherwise.

Assume the system is in some configuration \ls{g}. If there exist some
enabled nodes, a {\em daemon}\footnote{The daemon achieves the
  asynchrony of the system.} selects a non-empty set of them; every
chosen node {\em atomically} executes its algorithm, leading to a new
configuration \ls{g'}. The transition from \ls{g} to \ls{g'} is called
a {\em step}. To model steps in \Coq, we use functions with type
\lstinline{(Diff := Node $\to$ option State)}. We simply call
\emph{difference} a variable of type \ls{Diff}. A difference
contains the updated states of the nodes that actually execute some
action during the step, and maps any other node to $\bot$.
Steps are defined as a binary relation $\mapsto$ over configurations
expressed in \Coq by the relation \ls{Step}: \ls{(Step g' g)} holds
for \lstinline{g $\mapsto$ g'}.\footnote{Note the inverse
  order of the parameters in \ls{Step}.} It requires that there exists
a difference \ls{d} such that
\begin{itemize}
\item at least one node actually changes its state,,
\item every update in \ls{d} corresponds to the execution of the
  algorithm, namely, \ls{run}
\item and the next configuration,
  \ls{g'}, is obtained applying the function \ls{(diff_eval d g)}
  given by:
  \lstinline{$\forall$(p: Node), (g' p) = (d p)} if \lstinline{(d p) $\neq$ $\bot$}, and \ls{(g' p) = (g p)} otherwise.
\end{itemize}
An \emph{execution} of $\Algo$ is a sequence of configurations
\lstinline{g$_0$ g$_1$ $\ldots$ g$_i$ $\ldots$} such that
\lstinline{g$_{i-1}$ $\mapsto$ g$_i$} for all $i>0$. Executions
may be finite or infinite and are modeled in \Coq with the type
\begin{lstlisting}
CoInductive Exec: Type :=
| e_one:  Env $\to$ Exec
| e_cons: Env $\to$ Exec $\to$ Exec
\end{lstlisting}
and the predicate
\begin{lstlisting}
CoInductive is_exec: Exec -> Prop :=
| i_one:  $\forall$(g: Env), terminal g $\to$ is_exec (e_one g)
| i_cons: $\forall$(e: Exec) (g: Env),
            is_exec e $\to$ Step (Fst e) g $\to$ is_exec (e_cons g e)
\end{lstlisting}
where the keyword \ls{CoInductive} generates a greatest fixed point
capturing potentially infinite constructions\footnote{As opposed to
  this, the keyword \ls{Inductive} only captures finite
  constructions.}.
Considering first the constructor \ls{i_cons}, function \ls{(Fst e)}
returns the first configuration of execution \ls{e}. Thus, a variable
\ls{(e: Exec)} actually represents an execution of $\Algo$ when
\ls{(is_exec e)} holds, since each pair of consecutive configurations
\ls{g}, \ls{g'} in \ls{e} satisfies \ls{(Step g' g)}.

Considering now the constructor \ls{i_one}, proposition \ls{(terminal
  g)} means that \ls{g} is a \emph{terminal} configuration, namely, no
action of ${\Algo}$ is enabled at any node in \ls{g}. In our
framework, a terminal configuration is any configuration \ls{g} where
\ls{run} returns $\bot$ {\em for every node}.  Note that this
predicate is decidable thanks to the emptiness test. The predicate
\ls{is_exec} requires that only executions that end with a terminal
configuration are finite; every other execution is infinite.

As previously stated, each step from a configuration to another is
driven by a \emph{daemon}.  In our case study, we assume that the
daemon is {\em distributed} and {\em unfair}. \emph{Distributed} means
that while the configuration is not terminal, the daemon should select
at least one enabled node, maybe more. \emph{Unfair} means that there
is no fairness constraint, {\it i.e.}, the daemon might never select
an enabled node unless it is the only one enabled. Notice that the
propositions \ls{Step} and \ls{is_exec} are sufficient to handle the
distributed unfair daemon.

\subsubsection{Read-Only Variables}


We allow a part of a node state to be read-only: this is modeled with
the type \ls{ROState} and by the function %
\lstinline{(RO_part: State $\to$ ROState)} which typically represents
a subset of the variables handled in the \ls{State} of the node. The
projection \ls{RO_part} is extended to configurations by the
function \ls{(ROEnv_part g :=} \ls{(fun (p: Node)} \ls{=> RO_part (g p)))},
which returns a value of type \ls{ROEnv}.

We add the property \ls{RO_stable} to express the fact that those
variables are actually read-only, namely no execution of
\ls{run} can change their values.
From the assumption \ls{RO_stable}, we show that any property defined
on the read-only variables of a configuration is indeed preserved
during steps.

The introduction of Read-Only variables has been motivated by the fact
that we want to encompass the diversity of anonymity levels from the
distributing computing literature, {\it e.g.}, fully anonymous,
semi-anonymous, rooted, fully identified networks, {\em etc}.  By
default, our \Coq model defines fully anonymous network thanks to the
distinction between nodes (type \ls{Node}) and channels (type
\ls{Channel}).  We enriched our model to reflect other assumptions.

For example, consider the fully identified assumption. Identifiers are
typically constant data, stored in the node states. In our model, they
would be stored in the read-only part of the state.  Furthermore,
identifiers should be constant and unique all along the execution of
the algorithm (see the assumption in the case study). This means they
should be unique in the initial configuration and kept constant during the
whole execution.

We define a predicate \ls{Assume_RO} on \ls{ROEnv} (in the case of
fully identified assumption, \ls{Assume_RO} would express uniqueness
of identifiers) that will be assumed in each initial
configuration. From \ls{RO_stable}, this property will remain true all
along any execution.
Furthermore, the predicate \ls{Assume_RO} can express other
assumptions on the network such as connected networks or tree networks
(for this latter, see again the case study). As a shortcut, for any
configuration \ls{(g: Env)}, we use notation \ls{Assume g :=}
\ls{(Assume_RO} \ls{(ROEnv_part g))}.

\subsection{Setoids}
\label{sec:setoids}

When using \Coq function types to represent configurations and
differences, we need to state pointwise function equality, which
equates functions having equal values (extensional equality).  The \Coq
default equality is inadequate for functions since it asserts equality
of implementations (intensional equality). So, instead we chose to use
the setoid paradigm: we endow every base type with an
\emph{equivalence relation}. Setoids are commonly used in \Coq for
subsets, function sets, and to represent set-theoretic quotient sets
(such as rational numbers or real numbers); in particular we make use
of libraries \ls{Coq.Setoids.Setoid} and \ls{Coq.Lists.SetoidList}.

Consequently, every function type is endowed with a \emph{partial
  equivalence relation} (\textit{i.e.}, symmetric and transitive)
which states that, given equivalent inputs, the outputs of two
equivalent functions are equivalent. However, we
  also need reflexivity to reason about it, {\em i.e.}, functions are
  equivalent to themselves. In the context of partial equivalence
  relations, objects that are equal to themselves are said to be
  \emph{proper} elements (in \Coq : \lstinline{Proper R x := R x x}).
  For example, all elements of base types are proper since we use
  equivalence relations. Proper functions are also called
  \emph{compatible} functions or \emph{relation morphisms}: they
  return equivalent results when executed with equivalent parameters.
  Through all the framework, we assume \emph{compatible configurations
    and differences only}.  We also prove compatibility (properness)
  for every function and predicate defined in the sequel.
Additionally, we assume that equivalence relations on base types are
decidable.

As an example, let us consider configurations. The equality for type
\ls{Node} is noted \ls{(eqN: relation Node)} and assumes %
\begin{lstlisting}
  (eqN_equiv: Equivalence eqN;  eqN_dec: Decider eqN).
\end{lstlisting}
Note that \ls{(relation Node)} stands for %
\lstinline{(Node $\to$ Node $\to$ Prop)}; \ls{(Equivalence eqN)}
defines the conjunction of reflexivity, symmetry, and transitivity of
the relation \ls{eqN}; \ls{(Decider eqN)} expresses that the relation
is decidable by %
\lstinline!$\forall$(p p': Node), {eqN p p'} + {$\lnot$ eqN p p'}!,
where \ls{\{A\} + \{B\}} is the standard \Coq notation for
computational disjunction between \ls{A} and \ls{B}, \textit{i.e.},
Booleans carrying proofs of \ls{A} or \ls{B}.

The decidable equivalence relation on type \ls{State}, noted \ls{eqS} is 
defined similarly.
Now, the equality between configurations, which are functions of type
\lstinline{(Env: Node $\to$ State)}, is defined by %
\lstinline{eqE := (eqN ==> eqS).}%
This means that, for any two configurations \ls{(g1 g2: Env)},
\ls{(eqE g1 g2)} is defined by
\begin{lstlisting}
   $\forall$(p1 p2: Node), eqN p1 p2 $\to$ eqS (g1 p1) (g2 p2)
\end{lstlisting}

(in this model, \ls{(eqN p1 p2)} means that \ls{p1} and \ls{p2}
represent the same node in the network). Note that \ls{eqE} is not
reflexive \textit{a priori}. 
We enforce reflexivity assuming
compatible configurations only: any compatible configuration
\ls{(g: Env)} will satisfy
\begin{lstlisting}{}
   Proper eqE g := $\forall$ (p1 p2: Node), eqN p1 p2 $\to$ eqS (g p1) (g p2)
\end{lstlisting}
This means that
for any two equivalent nodes \ls{p1} and \ls{p2}, \textit{i.e.}, such
that \ls{(eqN p1 p2)}, we expect that \ls{(g p1)} and \ls{(g p2)}
produce the same result with respect to \ls{eqS}: \ls{(eqS (g p1) (g
  p2))}.





\section{Self-Stabilization and Silence}
\label{sec:silent_self_stab}

In this section, we express self-stabilization~\cite{Dijkstra74} in
the locally shared memory model with composite atomicity using \Coq
properties.

\subsection{Self-Stabilization}

Consider a distributed algorithm $\Algo$.
Let $\dsS$ be a predicate on executions %
(type \lstinline{(Exec $\to$} \ls{Prop)}).
$\Algo$ is \emph{self-stabilizing w.r.t. specification $\dsS$} if there
exists a predicate $\dsP$ on configurations %
(type \lstinline{(Env $\to$ Prop)}) such that:
\begin{itemize}
\item
  $\dsP$ is {\em closed} under $\Algo$, \textit{i.e.}, for each
  possible step \lstinline{g $\mapsto$ g'}, \lstinline{($\dsP$ g)}
  implies \lstinline{($\dsP$ g')}:
  \begin{lstlisting}
closure $\dsP$ := $\forall$(g g': Env), Assume g $\to$ $\dsP$ g $\to$ Step g' g $\to$ $\dsP$ g'
  \end{lstlisting}
\item
  $\Algo$ {\em converges} to $\dsP$, \textit{i.e.}, every execution
  contains a configuration which satisfies $\dsP$: 
  \begin{lstlisting}
convergence $\dsP$ := $\forall$(e: Exec),
  Assume (Fst e) $\to$ is_exec e $\to$
  safe_suffix (fun suf: Exec => $\dsP$ (Fst suf)) e
  \end{lstlisting}
  where \ls{(safe_suffix S e)} inductively checks that execution
  \ls{e} contains a suffix that satisfies \ls{S}.
\item
  $\Algo$ \emph{meets} $\dsS$ from $\dsP$, \textit{i.e.}, every
  execution which starts from a configuration where $\dsP$ holds,
  satisfies $ \dsS$:
  \begin{lstlisting}
spec_ok $\dsS \; \dsP$ := $\forall$(e: Exec),
  Assume (Fst e) $\to$ is_exec e $\to$ $\dsP$ (Fst e) $\to$ $\dsS$ e.
  \end{lstlisting}
\end{itemize}
%
The configurations which satisfy the predicate $\dsP$ are called {\em
  legitimate configurations}. The following predicate
  characterizes the property of being self-stabilizing for an
  algorithm:
\begin{lstlisting}
  self_stab $\dsS$ := $\exists$$\dsP$, closure $\dsP$ $\wedge$ convergence $\dsP$ $\wedge$ spec_ok $\dsS$ $\dsP$.
\end{lstlisting}

\subsection{Silence}

An algorithm is \emph{silent} if the communication between the
nodes is fixed from some point of the execution~\cite{DolevGS96}.
This latter definition can be transposed in the locally shared memory
model as follows: $\Algo$ is \emph{silent} if all its executions are
finite.
\begin{lstlisting}
Inductive finite_exec: Exec $\to$ Prop :=
| f_one: $\forall$(g: Env), finite_exec (e_one g)
| f_cons: $\forall$(e: Exec) (g: Env),
              finite_exec e $\to$ finite_exec (e_cons g e).
silence := $\forall$(e: Exec), Assume (Fst e) $\to$ is_exec e $\to$ finite_exec e.
\end{lstlisting}

By definition, executions of a \emph{silent and self-stabilizing}
algorithm \textit{w.r.t} some specification $\dsS$ end in
configurations which are usually used as legitimate configurations,
{\em i.e.}, satisfying $\dsP$. In this case, $\dsS$ can only allow
constrained executions made of a single configuration which is
legitimate; $\dsS$ is then noted $\dsS_\dsP$. To prove that $\Algo$ is
both silent and self-stabilizing {\em w.r.t.} $\dsS_\dsP$, we use, as
commonly done, a sufficient condition which requires to prove that
\begin{itemize}
\item
  all terminal configurations of $\Algo$ satisfy $\dsP$:
  \begin{lstlisting}
  P_correctness $\dsP$ := $\forall$(g: Env), Assume g $\to$ terminal g $\to$ $\dsP$ g
  \end{lstlisting}
\item
  and all executions of $\Algo$ are finite:
  \begin{lstlisting}
  termination := $\forall$(g: Env), Assume g $\to$ Acc Step g.
  \end{lstlisting}
\end{itemize}
The latter property is expressed with \ls{(Acc Step g)} for
every configuration \ls{g}.  The inductive proposition \ls{Acc} is
taken from Library \ls{Coq.Init.Wf} which provides tools on
well-founded induction. The accessibility predicate \ls{(Acc Step
  g)} is translated into
\begin{lstlisting}
  ($\forall$(g': Env), Step g' g $\to$ Acc Step g') $\to$ Acc Step g
\end{lstlisting}
Namely, the base case of induction holds when no step is possible from
current configuration \ls{g} and then, inductively, any configuration
\ls{g'} that eventually
 reaches such a terminal configuration
satisfies \ls{(Acc Step g')}. 
%

The sufficient condition, used to prove that an algorithm is both silent and
self-stabilizing, is expressed and proven by:
\begin{lstlisting}
Lemma silent_self_stab ($\dsP$: Env $\to$ Prop):
  P_correctness $\dsP$ $\wedge$ termination $\to$ silence $\wedge$ self_stab $\dsS_\dsP$.
\end{lstlisting}


\section{Tools for Proving Termination}
\label{sec:proof-convergence-part-general}

Usual termination proofs are based on some global potential built from
local ones. For example, local potentials can be integers and
the global potential can be the sum of them. In this case, the
argument for termination may be, for example, the fact that the global
potential is lower bounded and strictly decreases at each step of the
algorithm.  Global potential decrease is due to the modification of
local states at some nodes, however studying aggregators such as sums
may hide scenarios, making the proof more complex. 
Instead, we build here a global potential as the multiset containing
the local potential of each node and
provide a sufficient condition for termination on this multiset.  Our
method is based on two criteria that are sufficient to meet the
conditions for using the Dershowitz-Manna well-founded ordering on
multisets~\cite{DershowitzM79}. Given those criteria, we can show
that the multiset of (local) potentials globally decreases at each
step.
Note that instead of developing our own library, we have built specialized termination theorems on top of existing work, namely the Coq Standard Library for the lexicographic product ordering and the \ls{CoLoR} library~\cite{color} for multisets and the Dershowitz-Manna ordering.

We also provide tools for algorithms that have (local) priorities on
actions, {\em e.g}, an action $i$ is enabled at node $p$ only if every
action $j$, with $j<i$, is disabled at $p$. The overall idea is to
ease the proof by considering a given set of actions separately from
the others and to prove the termination of the algorithm assuming that
only this set of actions is executed.  Once termination is proved for
each set of actions separately, we use tuples of multisets ordered
with the lexicographical order to prove the termination of the whole
algorithm.

\subsection{Steps}
\label{sec:safeQstep}

One difficulty we faced, when trying to apply our method
straightforwardly, is that we cannot always define the local potential
function at a node without assuming some properties on its local
state, and so on the associated configuration.  Thus, we had to assume
the existence of some stable set of configurations in which the local
potential function can be defined.  When necessary, we use our
technique to prove termination of a subrelation of the relation
\ls{Step}, provided that the algorithm has been initialized in the
required stable set of configurations.
This point is modeled by a predicate on configurations, %
\lstinline{(safe: Env $\to$ Prop)}, and a type %
\ls{safeEnv := \{ g | safe g \}} which represents the set of
\emph{safe configurations} into which we restrict the termination
proof. Precisely, \ls{safeEnv} is a type whose values are ordered
pairs containing a term \ls{g} and a proof of \ls{(safe g)}.  Safe
configurations should be stable, \textit{i.e.}, it is assumed that no
step can exit from the set using the proposition:
\begin{lstlisting}
  stable_safe := $\forall$(g g': Env), safe g $\to$ Step g' g $\to$ safe g'.
\end{lstlisting}
Steps of the algorithm for which termination will be proven is defined by
\begin{lstlisting}
  safeStep sg2 sg1 := Step (getEnv sg2) (getEnv sg1)
\end{lstlisting}
with \ls{(sg1 sg2: safeEnv)} two safe configurations and where
\ls{(getEnv sg1)} (resp. \ls{(getEnv} \ls{sg2)}) accesses the actual
configuration, of type \ls{Env}, of \ls{sg1} (resp. \ls{sg2}).  We aim
at proving that this relation is well-founded. Since we know that
property \ls{safe} is stable from \ls{stable_safe}, we have the
following lemma (proven by induction, starting from the assumption): %
\begin{lstlisting}
  Lemma Acc_Algo_Multiset: 
    well_founded safeStep $\to$ $\forall$(g: Env), safe g $\to$ Acc Step g.
\end{lstlisting}
Note that %
\lstinline{(well_founded R := $\forall$a, Acc R a)}, like \ls{Acc}, is
taken from the standard \Coq Library \ls{Coq.Init.Wf}. Hence, to prove
termination of the algorithm as defined in
Section~\ref{sec:silent_self_stab}, we prove that \ls{safeStep} is
well-founded and use the above lemma to guarantee that the whole
algorithm terminates when initiated from any safe configuration.

We also allow restrictions on the kind of steps that will be handled
in termination proofs because, in some cases, it is easier to
partition steps and to prove termination of some kind of steps
separately from the others. We introduce the predicate %
\lstinline{QTrans: safeEnv $\to$} \lstinline{safeEnv $\to$ Prop} for
that purpose. The relation for which termination will be proven is
then defined by %
\begin{lstlisting}
  safeQStep sg2 sg1 := safeStep sg2 sg1 $\wedge$ QTrans sg2 sg1
\end{lstlisting}
for any two safe configuration \ls{(sg1} \ls{sg2:} \ls{safeEnv)}.  Note that
proving termination for all safe steps just consists in applying the
method with \ls{QTrans} defined as a tautology.

\subsection{Potential}
\label{sec:potential}

We assume that within safe configurations, each node can be endowed
with a potential value obtained using function \ls{(pot: safeEnv} $\to$
\ls{Node} $\to$ \ls{Mnat)}.  Notice that \ls{Mnat} simply represents
natural numbers\footnote{Natural numbers cover many cases and we
  expect the same results when further extending to other types of
  potential.}  encoded using the type from Library
\ls{CoLoR.MultisetNat} \cite{color}; it is equipped with the usual
equivalence relation, noted $=_{\tt P}$, and the usual well-founded
order on natural numbers, noted $<_{\tt P}$.

\subsection{Multiset Ordering}
\label{sec:multiset-ordering}

We recall that a multiset of elements in the setoid $P$ endowed with
its equivalence relation $=_P$, is defined as a set containing
\emph{finite numbers of occurrences (w.r.t. $=_P$) of elements of
  $P$}. Such a multiset is usually formally defined as a multiplicity
function $m: P\rightharpoonup\mathbb{N}_{\geq 1}$ which maps any element
to its number of occurrences in the multiset. We focus here on
 \emph{finite multisets}, namely, multisets whose
 multiplicity function has finite support. 
%
We define equality between multisets, noted $\eqM$, as the equality
between multiplicity functions.
Now, we assume that $P$ is also ordered using relation $<_P$,
compatible with $=_P$.  We use the Dershowitz-Manna order on finite
multisets \cite{DershowitzM79} defined as follows: the multiset $N$ is
smaller than the multiset $M$, noted $N \ltM M$, if and only if there
are three multisets $X$, $Y$ and $Z$ such that
\begin{itemize}
\item $N$ is obtained from $M$ by removing all elements in $X$ and
  adding all elements in $Y$. Elements in $Z$ are present in both $M$
  and $N$, and '$+$' between multisets means adding multiplicities, namely
  $$M \eqM Z+X \wedge N \eqM Z+Y$$
\item at least one element is removed, \textit{i.e.},
  $$X \neqM \emptyset$$
\item each element that is added ({\em i.e.} in $Y$) must be smaller
  ({\em w.r.t.} $<_P$) than some removed element ({\em i.e.} in $X$),
  that is:
  $$\forall y \in Y, \exists x \in X, y <_{P} x$$
\end{itemize}
 It is
shown~\cite{DershowitzM79} that if $<_P$ is a well-founded order, then
the corresponding order $\prec$ is also well-founded.

In our context, we consider finite multisets over \ls{Mnat},
(\textit{i.e.}, $=_P$ is $=_{\tt P}$ and $<_P$ stands for $<_{\tt P}$).
We have chosen to model them as lists of elements of \ls{Mnat} and we
build the potential of a configuration as the multiset of the
potentials of all nodes, {\em i.e.}, a multiset of (local) potentials of a
configuration \ls{(sg: safeEnv)} is defined by
\begin{lstlisting}[basicstyle=\tt\small]
  Pot sg := List.map (pot sg) all_nodes
\end{lstlisting}
where \ls{all_nodes} is the list of all nodes in the network (see
Section~\ref{sec:model}) and \ls{(List.map} \ls{f l)} is the standard
operation that returns the list of values obtained by applying \ls{f}
to each element of \ls{l}.
The corresponding Dershowitz-Manna order is defined using the library
CoLoR \cite{color}.  The library also contains the proof that %
\lstinline{(well_founded $<_{\tt P}$) $\to$} %
\lstinline{(well_founded $\prec$)}. %

%
%
Using this latter result and the standard result which proves
\lstinline{(well_founded $<_{\tt P}$)}, we easily deduce
\lstinline{(well_founded $\prec$)}.

%

\subsection{Termination Theorem}
\label{sec:termination-theorem}

Proving the termination of a set of safe steps then consists in
showing that for any such a step, the corresponding global potential
decreases {\em w.r.t.} the Dershowitz-Manna order $\ltM$. We call this
proof goal \textit{safe inclusion}:
\begin{lstlisting}
  safe_incl := $\forall$(sg1 sg2: safeEnv), 
               safeQStep sg2 sg1 $\to$ (Pot sg2) $\ltM$ (Pot sg1).
\end{lstlisting}
We establish a sufficient condition made of two criteria on node
potentials which validates \ls{safe_incl}.
The \emph{Local Criterion} finds for any node \ls{p} whose potential has
increased, a witness node \ls{p'} whose potential has decreased from a
value that is even higher than the new potential of \ls{p}:
\begin{lstlisting}
  Hypothesis local_crit: $\forall$(sg1 sg2: safeEnv), safeQStep sg2 sg1 $\to$
    $\forall$(p: Node),   (pot sg1 p)  $<_{\tt P}$ (pot sg2 p) $\to$
    $\exists$(p': Node), (pot sg1 p') $\neq_{\tt P}$ (pot sg2 p') $\wedge$ 
                 (pot sg2 p)  $<_{\tt P}$ (pot sg1 p').
\end{lstlisting}
The \emph{Global Criterion} exhibits, at any step, a node whose potential
has changed:
\begin{lstlisting}
  Hypothesis global_crit: $\forall$(sg1 sg2: safeEnv), safeQStep sg2 sg1 $\to$
    $\exists$(p: Node), (pot sg2 p) $\neq_{\tt P}$ (pot sg1 p).
\end{lstlisting}
Assuming both hypotheses (see Section~\ref{sec:term-algocl} for the
instantiation of these
criteria), 
we are able to prove \ls{safe_incl} as follows: we define $Z$ as the
multiset of local
potentials 
that did not change, and $X$ (resp. $Y$) as the complement of $Z$ in
the multiset of local potentials \ls{(Pot sg1)} (resp.  \ls{(Pot
  sg2)}). Global criterion is used to show that $X\neq\emptyset$, and
local criterion is used to show that $\forall y \in Y, \exists x \in
X, y <_{\tt P} x$.
Since any relation included in a well-founded order is also
well-founded, we get that relation \ls{safeQStep} is well-founded.

\subsection{Lexicographical Order}
\label{sec:lexical-order}

We now provide tools to divide a termination proof according to given
subsets of safe steps. Usually, algorithms made of several actions
enforce priority between them. For example, actions can be prioritized
so that only one action is enabled at a given node at one time;
namely, the second action can only be enabled at a node if the first
action is disabled, and so on. In such a case, it is often more
convenient to consider each action separately, \textit{i.e.}, show
that when nodes execute a particular action only, the algorithm
converges and then generalize by gradually incorporating the other
ones. To that goal, we consider a partition of safe steps; we detail
here the case for two subsets, one having priority over the other.

We consider two relations over safe configurations, noted %
\lstinline{(Trans1 Trans2: safeEnv $\to$} %
\lstinline{safeEnv $\to$ Prop)}. This may represent steps induced by
two different actions. We assume for each a local potential %
\lstinline{(pot1 pot2: safeEnv $\to$ Node $\to$ Mnat)} and as in
Section~\ref{sec:multiset-ordering}, we build the corresponding
multisets of local potentials, for any safe configuration \ls{sg},
by %
\lstinline{(Pot$i$ sg :=} \lstinline{List.map (pot$i$ sg)}
\lstinline{all_nodes)} with $i\in\{1, 2\}$. Here we expect that
\ls{Trans1} has priority on \ls{Trans2}. To encode this priority, we
require that when a step from \ls{Trans2} occurs, the multisets of
potentials measured by the action from \ls{Trans1} (\ls{Pot1}) is left
unchanged:
\begin{lstlisting}
  Hdisjoint := $\forall$(sg sg': safeEnv), 
               Trans2 sg' sg $\to$ (Pot1 sg) $\eqM$ (Pot1 sg').
\end{lstlisting}

The idea is to prove that steps from \ls{Trans1} and \ls{Trans2} taken
together, namely the union of the relations \ls{Trans1} and
\ls{Trans2}, converge, provided that steps from \ls{Trans1}
(resp. \ls{Trans2}) terminate when taken separately.

We use the following ordering relation: for any two safe
configurations \ls{(sg sg': safeEnv)}, %
\lstinline{sg $<_{lex}$ sg'} is defined by
\begin{lstlisting}
    (Pot1 sg) $\ltM$ (Pot1 sg') 
  $\vee$ (Pot1 sg) $\eqM$ (Pot1 sg) $\wedge$ (Pot2 sg) $\ltM$ (Pot2 sg').
\end{lstlisting}
$<_{lex}$ is built using the lexicographical order from the Library
ColoR, applied on pairs of multisets of local potentials. We also use
results from this library to show that $<_{lex}$ is well founded, as
far as the order on local potentials, $<_{\tt P}$, is.

Now, the argument is the same as for the Termination Theorem (see
\ref{sec:termination-theorem}): any order included in a well-founded
order is also well-founded. We aim at showing that %
\lstinline{(Trans1} \lstinline{$\cup$} \lstinline{Trans2)} is well-founded using the following
argument: \lstinline{(Trans1} \lstinline{$\cup$} \lstinline{Trans2)} should be included into
relation $<_{lex}$.  To obtain this, we first use the assumption
\ls{Hdisjoint} that ensures priorities between \ls{Trans1} and
\ls{Trans2}. Second, we require safe inclusion for both relations
\ls{Trans1} and \ls{Trans2}, namely:
\begin{lstlisting}
  safe_incl1 := $\forall$(sg sg': safeEnv), 
                Trans1 sg' sg $\to$ (Pot1 sg') $\ltM$ (Pot1 sg).
  safe_incl2 := $\forall$(sg sg': safeEnv), 
                Trans2 sg' sg $\to$ (Pot2 sg') $\ltM$ (Pot2 sg).
\end{lstlisting}
This ensures that, at any safe step of \ls{Trans1}
(resp. \ls{Trans2}), the corresponding global potential
decreases. From this we obtain our goal:
\begin{lstlisting}
  Lemma union_lex_wf2: well_founded (Trans1 $\cup$ Trans2).
\end{lstlisting}
Our framework also contains the same results for three relations; the
lemma corresponding to the above one is called
\lstinline{union_lex_wf3}.

Proving the two assumptions are satisfied (namely, the
  priorities between \ls{Trans1} and \ls{Trans2}, and safe inclusion
  for both relations \ls{Trans1} and \ls{Trans2}) implies the
  termination of the algorithm, while considering both
  relations \ls{Trans1} and \ls{Trans2} separately. In particular, we
  can use the Termination Theorem given in \ref{sec:termination-theorem}
  to show the safe inclusion of each relation, by instantiating the
  local and global criteria for each relation. 



\section{Tools for Quantitative Properties}
\label{sec:proof-counting-part}

\label{sec:proof-counting-part-general}

To handle some quantitative properties of an algorithm, we have to set
up a library dealing with cardinality of sets in general and also
cardinals of {\em finite} sets. The need for a
  new library arises from the absence of setoid-compatible
  formalization of set cardinality. For example,
  the \lstinline{Ensembles} and \lstinline{Finite_Sets} module from
  the \Coq standard library relates subset (predicates) over a fixed
  universe type \lstinline{U}, and elements are considered up to
  Leibniz equality (see definition of \lstinline{Singleton}). Instead,
  we build a theory allowing to compare the cardinality of arbitrary
  setoids (\emph{i.e.} of sets of equivalence classes) built on top of
  distinct types instead of subsets over the same type.

The library contains basic properties about set
operations such as Cartesian product, disjoint union, and
subset. Proofs are conducted using standard techniques.

\subsection{Cardinality on Setoids}
\label{sec:cardinality-setoids}

To be able to order cardinalities, we define a property, called
\ls{Inj}, on a pair of setoids $(A, =_A)$ and $(B, =_B)$ which
requires the existence of an injective and compatible function,
\ls{inj}, from $A$ to $B$ whose domain is $A$. Namely:
\begin{itemize}
\item\ls{Inj_compat}: \ls{inj} is compatible (see
  Section~\ref{sec:setoids}),
\item\ls{Inj_left_total}: domain of \ls{inj} is $A$, \textit{i.e.}, any
  element in $A$ is related to at least one element in $B$,
\item\ls{Inj_left_unique}: \ls{inj} is injective, \textit{i.e.}, any
  element in $B$ is related to at most one (w.r.t. $=_A$) element in
  $A$.
\end{itemize}
Relation \ls{Inj} is proven reflexive and transitive. 
We model cardinality ordering using the three-valued type
\ls{(Card_Prop := Smaller | Same | Larger)} and the following property
\ls{Card}. \ls{Card} distinguishes the different ways \ls{Inj} can
apply to pairs of setoids:
\begin{itemize}
\item \lstinline{(Card Smaller $A$ $B$)}\footnote{We omit parameters
    $=_A$ and $=_B$ for better readability.} is defined by %
  \lstinline{(Inj $A$ $B$)} which expresses that $A$ has a cardinal
  smaller or equal to that of $B$, w.r.t. equalities $=_A$ and $=_B$;
\item Similarly, \lstinline{(Card Larger $A$ $B$)} is defined by
  \lstinline{(Inj $B$ $A$)}
\item and \lstinline{(Card Same $A$ $B$)} by %
\lstinline{(Inj $B$ $A$ $\wedge$ Inj $A$ $B$)}.
\end{itemize}
\ls{(Card prop)} is reflexive and transitive for any value of \ls{prop}
in \ls{Card_prop}. It is also antisymmetric in the sense that
\ls{(Card Smaller)} and \ls{(Card Larger)} implies \ls{(Card Same)}
for a given pair of setoids (trivial from the definitions).

\subsection{Finite Cardinalities}
\label{sec:finite-cardinalities}

We now focus on finite setoids and define tools to express their
cardinalities. We first define, for a given natural number $N$, the
setoid %
\begin{lstlisting}
  $\mathcal{M}_N$ := {$i$: nat | $i < N$}.
\end{lstlisting}
$\mathcal{M}_N$ simply models the set of natural numbers $\{0, 1, ...,
N - 1\}$.\footnote{In \Coq, the values of $\mathcal{M}_N$ are ordered
  pairs containing a natural number $i$ and a proof of %
  $i < N$ and $\mathcal{M}_N$ is equipped with the
  standard equality on type \ls{nat}, wrapped to be able to compare
  values of type $\mathcal{M}_N$.}
We first proved that \ls{Inj} captures finite cardinality ordering
\begin{lstlisting}
  Lemma Inj_le_iff: $\forall$($m$ $n$: nat), Inj $\mathcal{M}_m$ $\mathcal{M}_n$ $\leftrightarrow$ $m \leq n$.
\end{lstlisting}
and the corresponding corollaries with \ls{Card}, \emph{e.g.},
\begin{lstlisting}
  $\forall$($m$ $n$: nat), Card Smaller $\mathcal{M}_m$ $\mathcal{M}_n$ $\leftrightarrow$ $m \leq n$.
\end{lstlisting}
(similar corollaries exist for \ls{Larger} and \ls{Same}). The
following predicate \ls{Num_Card} is then used to express that a
setoid $A$ has cardinality at least (resp. at most, resp. equal to)
some natural number $n$ with %
\lstinline{Num_Card prop $A$ $n$ := (Card prop $A$ $\mathcal{M}_n$)}
where \ls{prop} is any \ls{Card_Prop}. For instance, %
\lstinline{(Num_Card Smaller $A$ $n$)} means that $A$ contains at most
$n$ elements w.r.t. $=_A$.

\subsection{Cartesian Products}
\label{sec:cartesian-products}

We developed results about Cartesian products. First, the Cartesian
product is monotonic {\em w.r.t.} cardinality:
\begin{lstlisting}
  Lemma Inj_prod: $\forall$prop, Card prop $A_1$ $A_2$ $\to$ Card prop $B_1$ $B_2$ $\to$ 
    Card prop ($A_1 \times B_1$) ($A_2 \times B_2$).
\end{lstlisting}
where $(A_1,=_{A_1})$, $(A_2,=_{A_2})$, $(B_1,=_{B_1})$,
$(B_2,=_{B_2})$ are any setoids. Now, we showed that:
\begin{lstlisting}
  $\forall$ $n$ $m$: nat, Card Same ($\mathcal{M}_n\times\mathcal{M}_m$) $\mathcal{M}_{n{\times}m}$
\end{lstlisting}
namely, the Cartesian product of $\mathcal{M}_n = \{0,..., n-1\}$ and
$\mathcal{M}_m = \{0,...,m-1\}$ contains the same number of elements
as $\mathcal{M}_{n{\times}m} = \{0,..., n{\times}m-1\}$. This latter result is shown
using encoding functions from $\mathcal{M}_n\times\mathcal{M}_m$ to
$\mathcal{M}_{n{\times}m}$ and from $\mathcal{M}_{n{\times}m}$ to
$\mathcal{M}_n\times\mathcal{M}_m$. This intermediate result allows one to
easily deduce that the cardinality of a Cartesian product is the
product of cardinalities:
\begin{lstlisting}
  $\forall$prop ($n$ $m$: nat), Num_Card prop $A$ $n$ $\to$ Num_Card prop $B$ $m$ $\to$ 
    Num_Card prop ($A\times B$) ($n{\times}m$)
\end{lstlisting}

\subsection{Disjoint Unions}
\label{sec:disjoint-unions}

We developed similar lemmas about the disjoint union of sets, noted
\ls{+}. The main results is:
\begin{lstlisting}
  $\forall$prop ($n$ $m$: nat), Num_Card prop $A$ $n$ $\to$ Num_Card prop $B$ $m$ $\to$ 
    Num_Card prop ($A + B$) ($n + m$)
\end{lstlisting}

\subsection{Subsets}
\label{sec:subsets}

We proved many toolbox results, about subsets, which are expressed
using \ls{Card} as well as \ls{Num_Card}. For instance,
\begin{itemize}
\item any subset of a set $A$ has \ls{Smaller} cardinality than that
  of $A$,
\item a set is one of its subsets with \ls{Same}
  cardinality,
\item the empty subset contains 0 element,
\item a non-empty set contains at least 1 element,
\item a singleton contains exactly one element.
\end{itemize}

\subsection{Number of Elements in Lists}
\label{sec:numb-elem-lists}

To prove the existence of finite cardinality for finite setoids, we
use lists, since, for example, the setoid of nodes of the network is
encoded in our framework as the list \ls{all_nodes}. We now consider a
setoid $A$, whose equality $=_A$ satisfies the classical excluded
middle property %
\lstinline{($\forall$a1 a2: $A$, a1 $=_A$ a2 $\vee$ a1 $\neq_A$ a2)}
and a predicate function \lstinline{(P: A $\to$ Prop)}, which also
satisfies the classical excluded middle property %
\lstinline{($\forall$a: $A$, P a $\vee$ $\lnot$ P a)}. Under these
conditions, we can prove:
\begin{lstlisting}
  $\forall$(l: list $A$), $\exists$($n$: nat), Num_Card Same {a: $A$ | P a $\wedge$ a $\in_{=_A}$ l} $n$
\end{lstlisting}
namely, for any list \ls{l}, the set of elements in \ls{l}
(w.r.t. $=_A$) which satisfies predicate \ls{P} has finite cardinality
$n$.
Or, equivalently, assuming the existence of a list \ls{l} which
contains every element of type $A$, we get that the number of elements
which satisfy \ls{P} is finite:
\begin{lstlisting}
  $\forall$(l: list $A$), ($\forall$(a: $A$), a $\in_{=_A}$ l) $\to$ 
      $\exists$($n$: nat), Num_Card Same {a: $A$ | P a } $n$.
\end{lstlisting}
When predicate function \ls{P} returns \ls{True} for all argument values,
this provides the number of elements of list \ls{l}, up to $=_A$.



\section{\texorpdfstring{$k$}{k}-Clustering Algorithm \texorpdfstring{$\algoCLR{k}$}{C(k)}}
\label{sec:k-clust-algor}

We have certified a non trivial part of the silent self-stabilizing
algorithm proposed in~\cite{DLDHR12c}. Given a non-negative integer
$k$, this algorithm builds a $k$-clustering of a bidirectional
connected network $G= (V,E)$ containing at most $\lfloor
\frac{n-1}{k+1} \rfloor + 1$ $k$-clusters, where $n$ is the number of
nodes.  A \emph{$k$-cluster} of $G$ is a set $C \subseteq V$, together
with a designated node $h \in C$, such that each member of $C$ is
within distance $k$ of $h$.\footnote{The distance $\|p,q\|$ between
  two nodes $p$ and $q$ is the length of a shortest path linking $p$
  to $q$ in $G$.}  A {\em $k$-clustering} is then a partition of $V$
into distinct $k$-clusters.

The algorithm proposed in \cite{DLDHR12c} is actually a hierarchical
collateral composition~\cite{DLDHR13} of two silent self-stabilizing
sub-algorithms: the former builds a rooted spanning tree, the latter
is a $k$-clustering construction which stabilizes once a rooted
spanning tree is available in the network.
In this paper, we focus in the certification of the second part,
namely, the construction, in a self-stabilizing and silent way, of a
$k$-clustering on a rooted spanning tree containing at most $\lfloor
\frac{n-1}{k+1} \rfloor + 1$ clusterheads. The $k$-clustering is
actually organized as a spanning forest. Each $k$-cluster is an
in-tree of height at most $k$ rooted at its clusterhead. Moreover,
each $k$-cluster is colored with the identifier of its
clusterhead. Hence, each node $p$ should compute the identifier of its
clusterhead $c$ and the channel corresponding to its {\em parent
  link}, that is, the link from $p$ to its {\em parent} in the
$k$-cluster, that is, the unique neighbor of $p$ on the shortest path
from $p$ to $c$ in the $k$-cluster.

{\small
\begin{algorithm}[pt]
  \caption{$\algocl{k}$, code for each {\nde} $p$
    \label{alg:clk}}
  \begin{minipage}[t]{0.98\linewidth}
    
  \smallskip

    {\bf Constant Input:\-}
    $\id(p) \in \Ids$; $\parent{p}$ $\in \neighbors{p} \cup \{\perp\}$
    ~\\

    {\bf Variable:\-}
    $p.\alpha$ $\in \Z$;
    $p.\parCl$ $\in 
\neighbors{p} \cup \{\perp\}$; $p.\head \in \Ids$ \\
    ~\\
{\bf Predicates:}\\
  \begin{tabular}{lcl}
    $\IsShort(p)$ & $\equiv$ & $p.\alpha < k$ \\
    $\IsTall(p)$ & $\equiv$ & $p.\alpha \geq k$ \\
    $\kDominator(p)$ & $\equiv$ & $(p.\alpha = k) \vee (\IsShort(p)
    \wedge \parent{p} = \perp)$ \\
~\\
\end{tabular}\\
  {\bf Macros:} \\
  \begin{tabular}{lcl}
    $\ShortChildren(p)$ & $=$ & $\{q\in \neighbors{p}\ |\ \parent{q} =
    \rho_p(q) \wedge \IsShort(q)\}$ \\
    $\TallChildren(p)$ & $=$ & $\{q\in \neighbors{p}\ |\ \parent{q} =
    \rho_p(q) \wedge \IsTall(q)\}$ \\
    ~\\
    $\MaxAShort(p)$ & $=$ & $\max (\{q.\alpha\ |\ q \in \ShortChildren(p)\} \cup \{-1\})$\\
    ~\\
    $\MinATall(p)$ & $=$ & $\min (\{q.\alpha\ |\ q \in \TallChildren(p)\} \cup \{2k+1\})$\\
    ~\\
    $\MinIDMinATall(p)$ & $=$ &
    {\bf if} $\TallChildren(p) = \emptyset$
    {\bf then} $\perp$\\
    &     & {\bf else} $\min_{<_C} {\{ q \in \TallChildren(p) ~|~ q.\alpha =
      \MinATall(p)\}}$ \\
    ~\\
    $\Alpha(p)$ & $=$ &
    {\bf if} $\MaxAShort(p) + \MinATall(p) \leq 2k-2$
    {\bf then} $\MinATall(p)+1$ \\
    & & {\bf else} $\MaxAShort(p)+1$ \\
    ~\\
    $\ParCLR(p)$ & $=$ &
    {\bf if} $p.\alpha = k$
    {\bf then}  $\perp$\\
    &     & {\bf else if} $IsShort(p)$ \textbf{then} $\parent{p}$
    \textbf{else} $\MinIDMinATall(p)$ \\
    ~\\
    $\HeadCLR(p)$ & $=$ &
    {\bf if} $\kDominator(p)$
    {\bf then}  $\id(p)$ {\bf else} \\
    &     & {\bf else if} $p.\parCl \notin \neighbors{p}$ \textbf{then} $p.\head$
    \textbf{else} $p.\parCl.\head$ \\
    ~\\
  \end{tabular}

  {\bf Actions:} \\
  \begin{tabular}[t]{lcl}
    $p.\alpha \neq \Alpha(p)\ $ & $\hookrightarrow\ $ & $p.\alpha
    \gets \Alpha(p)$ \\
    $p.\alpha = \Alpha(p) \wedge p.\parCl \neq \ParCLR(p)$
    & $\hookrightarrow$ & $p.\parCl \gets \ParCLR(p)$ \\
    $p.\alpha = \Alpha(p) \wedge p.\parCl = \ParCLR(p) \wedge p.\head
    \neq \HeadCLR(p)$
    & $\hookrightarrow$ & $p.\head \gets \HeadCLR(p)$ \\
  \end{tabular}
  \end{minipage}
\end{algorithm}
}


The code of the algorithm, called $\algocl{k}$, is given in
Algorithm~\ref{alg:clk}. We have used our framework to encode
$\algocl{k}$, its assumptions and specification, and to certify its
correctness.


\subsection{Assumptions}

\subsubsection{Unique Identifiers}

The algorithm $\algocl{k}$ requires that nodes are uniquely
identified: we assume a datatype for identifiers, noted $\Ids$, which
is endowed with a decidable equivalence relation noted \ls{eqId}. Each
node $p$ is equipped with a constant input $\id(p)$ of type $\Ids$
that represents its identifier.  We use the predicate \ls{uniqueId} to
represent uniqueness of the identifiers as follows:
\begin{lstlisting}
uniqueID Id := $\forall$(p1 p2: Node), eqId $\id$(p) $\id$(p2) -> eqN p1 p2
\end{lstlisting}

\subsubsection{Spanning Tree}

We denote the directed spanning tree and its root by $T$ and $\r$,
respectively: the knowledge of $T$ is locally distributed at each node
$p$ using the constant input $\parent{p}$ $\in \neighbors{p} \cup
\{\perp\}$. When $p \neq \r$, $\parent{p}$ $\in \neighbors{p}$ and
designates its parent in the tree (precisely, the channel outgoing to
its parent). Otherwise, $p$ is the root and $\parent{p} = \perp$.

We express the assumption about the spanning tree using predicate
\lstinline{(span_tree $\r$ Par)}. This predicate checks that the graph
$T$ induced by \ls{Par} is a subgraph of $G$ which actually encodes a
spanning tree rooted at $\r$ by the conjunction of
\begin{itemize}
\item $\r$ is the unique node such that $\parent{\r} = \perp$,
\item $\parent{p}$, for every non-root node $p$, is an existing
  channel outgoing from $p$,
\item $T$ contains no loop.
\end{itemize}
From the last point, we show that, since the number of nodes is
finite, the relation extracted from \ls{Par} between nodes and their
parents (resp. children) in $T$ is well-founded. We call this result
\ls{WF_par} (resp. \ls{WF_child}) and express it using
\ls{well_founded}.

\subsubsection{Predicate \ls{Assume}$_{cl}$}
We instantiate the predicate \ls{Assume_RO} to express that in
any configuration \ls{(g: Env)}, $G$ is bidirectional, identifiers are
unique, and a rooted spanning tree is available in $G$ ({\em n.b.},
this latter also implies that $G$ is connected):
\begin{lstlisting}
  Assume$_{cl}$ g := sym_net $\wedge$ uniqueId Id $\wedge$ $\exists$$\r$, span_tree $\r$ Par.
\end{lstlisting}

\subsection{Specification}
\label{sec:algo-spec}

The goal of algorithm $\algocl{k}$ is to compute a $k$-clustering
using the spanning tree $T$. We consider any positive parameter $k$,
(here, $k$ is taken in \ls{Z}, as for other numbers, and assumed to be
positive)
  and we model the $k$-clustering, \textit{i.e.}, the output of the
  algorithm, using the predicate \lstinline{kCluster}: for a given
  terminal configuration \ls{(g: Env)}, the proposition
  \lstinline{(kCluster g h p)} means that node $p$ is in the
  $k$-cluster of node \ls{h} and \ls{h} is a clusterhead; precisely,
  \ls{h} is the clusterhead of the $k$-cluster %
  \lstinline!{p | kCluster g h p}!.
Note that using this definition, the fact that \ls{h} is a
clusterhead is given by the predicate %
\lstinline{(clusterHead g h := kCluster g h h)}.
The predicate \ls{kCluster} actually designates a $k$-clustering when
\begin{itemize}
\item for any clusterhead $h$, the set %
  \lstinline!{ p | kCluster g h p }! actually represents a
  $k$-cluster, namely, for every node \ls{p} in this set, there exists
  a path in this $k$-cluster (\textit{i.e.}, the path is made of nodes
  \ls{q} such that \ls{kCluster g h q}) of length smaller than or
  equal to $k$ from \ls{p} to \ls{h}, and
\item the set of $k$-clusters is a partition of the set of nodes, or
  equivalently, every node belongs to a $k$-cluster and the
  intersection of any two distinct $k$-clusters is empty.
\end{itemize} 
The complete check for $k$-clustering is performed using the
conjunction of the two following predicates on configuration \ls{(g:
  Env)}:
\begin{lstlisting}
  kCluster_OK g := $\forall$(h: Node), clusterHead g h $\to$ 
    $\forall$(p: Node), kCluster g h p $\to$
      $\exists$(path: list Node), is_path h path p $\wedge$
                          ($\forall$(q: Node), q $\in$ path $\to$ kCluster g h q) $\wedge$ 
                          (length path) $\leq$ $k$
\end{lstlisting}

\label{partition_ok}
\begin{lstlisting}
  partition_OK g := $\forall$(p: Node), 
    ($\exists$(h: Node), kCluster g h p) $\wedge$
    ($\forall$(h h': Node), kCluster g h p $\to$ kCluster g h' p -> eqN h h')
\end{lstlisting}
where predicate \ls{is_path} detects if the list of nodes
\ls{path} actually represents a path in the network between the nodes
\ls{h} and \ls{p}, and \ls{length} computes the length of the path.

Actually, the algorithm computes a stronger specification. First, it
ensures that there are no more than $\lfloor \frac{n-1}{k+1} \rfloor +
1$ clusterheads in any terminal configuration \ls{(g: Env)}:
\begin{lstlisting}
  count_OK g := $(n - 1) \geq (k + 1) (|$CH$| - 1)$
\end{lstlisting}
where \ls{CH} is the set of clusterheads, given by %
\lstinline!{ h: Node | clusterHead g h }!.
Second, in a terminal configuration \ls{g}, each node knows the
identifier of its clusterhead and the channel corresponding to its
parent link in the $k$-cluster:
\begin{itemize}
\item there exists a local function (on state of a node), %
  \lstinline{(clusterHeadID: State $\to$ Ids)}, which provides the
  identifier of the clusterhead of the node, and
\item there exists a local function, %
  \lstinline{(clusterParent: State $\to$ Channel)}, which returns the
  channel outgoing to the parent of the node in the $k$-cluster.
\end{itemize}
This provides the third part of the specification, for a configuration
\ls{(g: Env)}:
\begin{lstlisting}
  kCluster_strong g := 
    $\forall$(h p: Node), kCluster g h p $\leftrightarrow$ eqId $\id$(h) (clusterHeadID (g p))
    $\bigwedge$
    $\forall$(h: Node), clusterHead g h $\to$
      $\forall$(p: Node), kCluster g h p $\to$
        $\exists$(path: list Node), agreed_cluster_path g h path p $\wedge$ 
                            (length path) $\leq$ $k$
\end{lstlisting}
where \ls{(agreed_cluster_path g h path p)} is true if
\begin{itemize}
\item \ls{(cluster_path g h path p)} holds, meaning that
  \ls{path} is a {\em cluster path} in the network linking \ls{h} to
  \ls{p}, \textit{i.e.}, the path described by the values of the
  \ls{clusterParent} pointers in \ls{g}, and
\item every node in \ls{path} declares the same \ls{clusterHeadID} in \ls{g}.
\end{itemize}
Note that for any configuration \ls{(g: Env)}, %
\ls{(kCluster_strong g)} enforces \ls{(kCluster_OK g)}. Hence, the
complete specification is given by the conjunction %
\label{kcluster_strong}
\begin{lstlisting}
  $\dsP_{cl}$ g := (kCluster_strong $\wedge$ partition_OK g $\wedge$ count_OK g)
\end{lstlisting}
for a configuration \ls{(g: Env)}.

\subsection{Algorithm \texorpdfstring{$\algocl{k}$}{C(k)} in \Coq}

We translate $\algocl{k}$ into the type
\ls{Algorithm}. First, the \emph{state} of each node $p$ contains,
in addition to $\id(p)$ and $\parent{p}$, 
\begin{itemize}
\item an integer variable $p.\alpha$,
\item a variable $p.\parCl$ which is either a channel or
  $\perp$, and
\item a variable $p.\head$ of type \Ids.
\end{itemize}
Hence, we have instantiated the \ls{State} of a node as a record
containing fields \ls{(Par:} \ls{option} \ls{Channel)}, \ls{(Id:
  Ids)}, \lstinline{($\alpha$: Z)}, \lstinline{($\head$: $\Ids$)} and
\lstinline{($\parCl$: option Channel)}. \ls{Par} and \ls{Id} are
declared as \emph{read-only} variables.

Note that to be able to compute \textit{a} path from each node to its
clusterhead, the algorithm requires that channels are totally ordered
(to be able to compute the minimum value on a set of channels). Hence
we assume a strict total order $<_C$ on the type \lstinline{Channel}.
Furthermore, we chose to encode every number in the algorithm as
integers in \ls{Z}, as $\alpha$ is, since some of them may be negative
(see $\MaxAShort$) and computations use minus (see $\Alpha$).

Now, \emph{every predicate and macro} of Algorithm~\ref{alg:clk} can
be directly encoded in \Coq: for a node \ls{p} and a current
configuration \ls{g}, mainly all of them depend on $\neighbors{\tt
  p}$, $\rho_{\tt p}$, \ls{(g p)}, and \ls{(local_env g p)}; then the
translation is quasi-syntactic (see Library \ls{KClustering_algo} in
the online browsing) and provides a definition of \ls{run}.
The definition of $\algocl{k}$, of type \ls{Algorithm}, comes with a
proof that \ls{run} is compatible, as a composition of compatible
functions, and also with a straightforward proof of \ls{RO_stable}
which asserts that the read-only parts of the state, \ls{Par} and
\ls{Id}, are constant during steps, when applying \ls{run}.

\subsection{Overview of \texorpdfstring{$\algocl{k}$}{C(k)}}

 A {\em $k$-hop dominating set} of a graph is a subset
  $\clrdom$ of nodes such that every node of the graph is within
  distance $k$ from at least one node of $\clrdom$. The $k$-clustering
  problem is related to the notion of $k$-hop dominating set, since
  the set of clusterheads of any $k$-clustering is, by definition, a
  $k$-hop dominating set.

Algorithm $\algocl{k}$ builds a $k$-clustering in two phases.  During
the first phase, $\algocl{k}$ computes the set of clusterheads as a
$k$-hop dominating set of the spanning tree $T$ (and so of $G$), using
the variables $\alpha$ and the first action. 
The second phase consists of building (using variables $\parCl$ and
$\head$, and the two other actions) a spanning forest : Algorithm
$\algocl{k}$ computes each $k$-cluster as an in-tree of height at most
$k$ rooted at one of the already computed clusterhead. Moreover, each
$k$-cluster will be colored with the identifier of its clusterhead.

\subsubsection{Building $\clrdom$}  $\clrdom$ is constructed in a
bottom-up fashion starting from the leaves of $T$, using the values of
$p.\alpha$ for all $p$.
Precisely, $\clrdom$ is defined as the set of nodes $p$ such that the
predicate $\kDominator(p)$ holds, namely, when $p.\alpha = k$, or
$p.\alpha < k$ and $p = r$ %
(\emph{i.e.}, $p$ is the root). %
The goal of variable $p.\alpha$ at each
node $p$ is twofold.  First, it allows to determine a path of length
at most $k$ from $p$ to a particular node $q$ of $\clrdom$ which acts
as a {\em witness} for guaranteeing the $k$-hop domination of
$\clrdom$. Consequently, $q$ will be denoted as $W\!itness(p)$ in the
following.  Second, once correctly evaluated, the value $p.\alpha$ is
equal to $\|p,x\|$, where $x$ is the furthest node in $T(p)$, the
subtree of $T$ rooted at $p$, that has the same witness as $p$.

The algorithm divides processes into {\em short} and {\em tall}
according to the value of their $\alpha$-variable: if $p$ satisfies
$\IsShort(p)$, {\em i.e.}, $p.\alpha < k$, then $p$ is said to be {\em
  short}; otherwise, $p$ satisfies $\IsTall(p)$ and is said to be {\em
  tall}.
In a terminal configuration, the meaning of $p.\alpha$ depends on
whether $p$ is {\em short} or {\em tall}.


\emph{If $p$ is short}, we have two cases: $p\neq \r$ or $p = \r$.
In the former case, $W\!itness(p) \in \clrdom$ is outside of $T(p)$,
that is, the path from $p$ to $W\!itness(p)$ goes through the parent
link of $p$ in the tree, and the distance from $p$ to $W\!itness(p)$ is
at most $k-p.\alpha$. See, for example, in Configuration (I) of Figure
\ref{fig:clr}, $k=2$ and $m.\alpha = 0$ mean that $W\!itness(m)$ is at
most at distance $k-0 = 2$, now its witness $g$ is at distance 2.

In the latter case, $p$ ($= \r$) may not be $k$-hop dominated by any
 {\nde} of $\clrdom$ inside its subtree and, by definition, there
is no {\nde} outside its subtree, indeed $T(p) = T$, see the root $a$
in Configuration (I) of Figure \ref{fig:clr}. Thus, $p$ must be placed
in $\clrdom$.

\emph{If $p$ is tall}, there is at least one {\nde} $q$ at
$p.\alpha-k$ hops below $p$ such that $q.\alpha =k$. Any such a
process $q$ belongs to $\clrdom$ and $k$-hop dominates $p$. Hence, $p$
can select any of them as {\em witness} (in the algorithm, we break
ties using the order on channels).


The path from $p$ to $W\!itness(p)$ goes through a tall child with
minimum $\alpha$-value.  See, for example, in Configuration (II) of
Figure \ref{fig:clr}, $k = 2$ and $a.\alpha = 4$ mean that
$W\!itness(a)$, here $c$, is $4-k = 2$ hops below $a$. In
Configuration (I), remark that there are two possible witnesses for
$c$ ($c.\alpha = 4$): $g$ and $h$, both are $c.\alpha-k = 2$ hops
below $c$.

Note that, if $p.\alpha = k$, then $p.\alpha-k=0$, that is, $p = q =
W\!itness(p)$ and $p$ belongs to $\clrdom$.

\subsubsection{Constructing the $k$-Clustering} The second phase of
$\algocl{k}$ partitions the nodes into distinct $k$-clusters, each of
which contains one clusterhead. Each $k$-cluster is actually built as
a {\em $k$-cluster spanning tree}, a tree containing all the nodes of
that $k$-cluster. Each $k$-cluster spanning tree is a subgraph of $T$
rooted at the clusterhead, possibly with the directions of some edges
reversed. Furthermore, the height of the $k$-cluster spanning tree is
at most $k$.

Each node $p$ of $\clrdom$ designates itself as clusterhead by setting
$p.\parCl$ and $p.\head$ to $\perp$ and $\id(p)$, respectively (see
the second and third actions in Algorithm~\ref{alg:clk}).

Other nodes $q$ designate their parent in the $k$-cluster, $q.\parCl$,
using second action as follows: (1) if $q$ is {\em short}, then
$q.\parCl$ should be its parent in the tree $T$; (2) if $q$ is {\em
  tall}, then $q$ selects $q.\parCl$ as its {\em tall} child in the
tree of minimum $\alpha$ value; as explained before we use channel
order to break ties (see $\MinIDMinATall(q)$ and $\min_{<_C}$ which
selects the minimum channel in a set).

Finally, identifiers of clusterheads are propagated in the $\head$
variables top-down in the $k$-clusters using the third action (see
macro $\HeadCLR$).


\subsubsection{Examples}

\begin{figure}
  \centering
  \includegraphics[width=\textwidth]{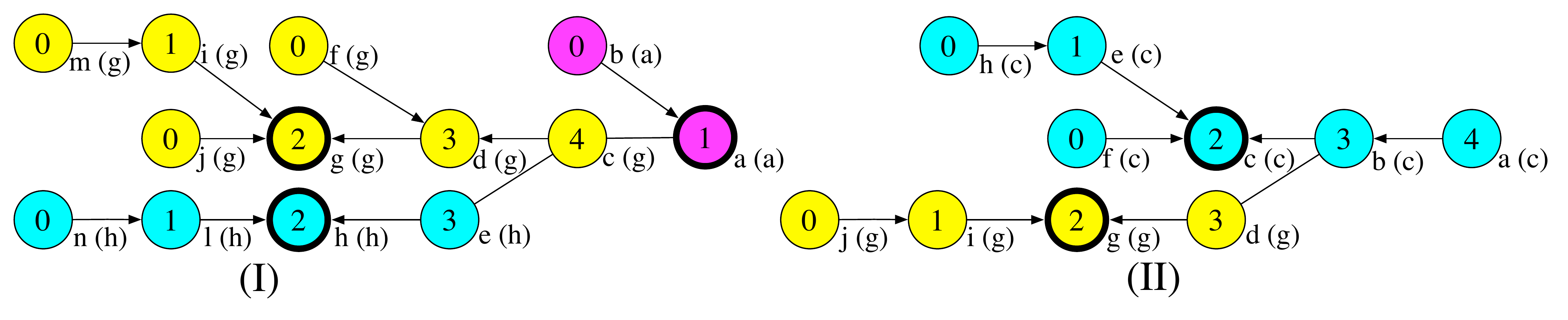}
  \caption{Two examples of $2$-clustering computed by $\algocl{2}$.
    We only draw the spanning tree, other edges are omitted. The root
    of each tree is the rightmost node. $\alpha$-values are given
    inside the nodes. Bold circles represent clusterheads.
    Identifiers are given next to the nodes. $\head$-values are given
    in brackets. Arrows represent $\parCl$ pointers. We colored the
    different $k$-clusters.}
  \label{fig:clr}
\end{figure}

Two examples of $2$-clustering computed by $\algocl{2}$ are given in Figure
\ref{fig:clr}.  In Subfigure \ref{fig:clr}.(I), the root is a {\em
  short} {\nde}, consequently it belongs to $\clrdom$. In Subfigure
\ref{fig:clr}.(II), the root is a {\em tall} {\nde}, consequently it
does not belong to $\clrdom$.

\section{Termination of \texorpdfstring{$\algoCLR{k}$}{C(k)}}
\label{sec:term-algocl}


The goal of this section is to build a certified proof of the
\ls{termination} property of $\algocl{k}$.  Namely, we show
that $\algocl{k}$ converges, as far as
\lstinline{Assume$_{cl}$} is satisfied.

Algorithm $\algocl{k}$ is made of three actions which are locally
prioritized at each node (see the guard of every action). The first
action (which computes $\alpha$) has priority over the second and
last actions. The second action (which computes $\parCl$) can only be
enabled when the first action is disabled, but has priority over the
last one. This latter action can be enabled only if the two others are
disabled. We use the method given in Section~\ref{sec:lexical-order}
to show the termination of $\algocl{k}$: it consists in proving the
termination of the three actions separately (using results from
Section~\ref{sec:termination-theorem}). Then, we use the lexicographic
order result from Section~\ref{sec:lexical-order} to conclude.

First, we assume \ls{sym_net} and a root node $\r$. For
the definition of safe configuration, we instantiate
\ls{safe} as every configuration in which read-only \ls{Par}-variables
satisfy: \lstinline{(span_tree $\r$ Par)}.  This assumption on the
existence of the spanning tree $T$ rooted at $\r$ is mandatory, since,
as we will see below, the local potentials we use in proofs are based
on the tree $T$.  Note that it is easy to prove that \ls{safe} is
stable, since it only depends on read-only variables.

\subsection{$\alpha$\lstinline{_SafeStep}}

For the first action, we build safe steps, called
\lstinline{$\alpha$_SafeStep}, by instantiating the predicate
\lstinline{QTrans} (see Section~\ref{sec:safeQstep}) as follows: for
any two safe configurations \ls{(sg sg': safeEnv)}, an
\lstinline{$\alpha$_SafeStep} occurs between \ls{sg} and \ls{sg'} if
\ls{(Step sg' sg)} holds and if the following condition is satisfied:
\begin{lstlisting}
  QTrans$_{\alpha}$ sg' sg := $\exists$(p: Node), $\alpha$_enabled sg p $\wedge$ has_moved sg' sg p
\end{lstlisting}
where \lstinline{$\alpha$_enabled} is exactly the guard of the first
action and \lstinline{(has_moved sg' sg p)} means that \ls{p} has
executed its local program during the step (this is done by checking
that the state of node \ls{p} in safe configuration \ls{sg'} is the
same as the result of \ls{run} on \ls{sg} and \ls{p}). We manage to
have 
Boolean versions of the above predicates; this is made possible due to
the fact that all nodes in the network are stored in the list
\ls{all_nodes}.
When \lstinline{($\alpha$_enabled sg p)}, we say that node \ls{p} is
{\em $\alpha$-enabled in the safe configuration \ls{sg}} and when %
\lstinline{(has_moved sg sg' p)} holds additionally, we say that
\ls{p} {\em has $\alpha$-moved from \ls{sg} to \ls{sg'}}.

In an \lstinline{$\alpha$_SafeStep}, we require that at least a node
executes its first action; note that this gives no guarantee on other
nodes, which may or may not execute an action.
We use then the method explained in
Section~\ref{sec:termination-theorem} to prove the safe inclusion of
\lstinline{$\alpha$_SafeStep}:
\begin{lstlisting}
Theorem $\alpha$_safe_inclusion:
  $\forall$(sg1 sg2: safeEnv), $\alpha$_SafeStep sg2 sg1 $\to$ (Pot$_{\alpha}$ sg2) $\ltM$ (Pot$_{\alpha}$ sg1).
\end{lstlisting}
In other words, when an \lstinline{$\alpha$_SafeStep} occurs, the global potential
\lstinline{Pot$_\alpha$} decreases. \lstinline{Pot$_\alpha$} is built
from $\alpha$, as the list of local potentials,
\lstinline{$\alpha$_pot}, at every node. In the following, we explain
how we compute this $\alpha$-potential, \lstinline{$\alpha$_pot}, at
each node.

\subsubsection{$\alpha$-Potential}

We define the \emph{depth} of a node as one plus the distance from the
root $\r$ to the node in the tree $T$. For a given safe configuration
\ls{sg} and a node \ls{p}, \ls{(depth sg p)} returns 1 (natural
number, type \ls{nat}) if \ls{p} is the root $\r$ and, otherwise,
\ls{(1 + (depth sg q))} where \ls{q} the parent of \ls{p} in the tree
$T$; the definition relies on structural induction on \ls{(WF_par p)}.
%
%
We define the $\alpha$-potential of a node \ls{p} in a safe
configuration \ls{sg}, \lstinline{($\alpha$_pot sg p)}, as 0 if \ls{p}
is not $\alpha$-enabled in \ls{sg} and \ls{(depth sg p)}, otherwise.

\subsubsection{Local Criterion for $\alpha$\lstinline{_SafeStep}s} 

Let \ls{sg1} and \ls{sg2} be two safe configurations where \lstinline{($\alpha$_safeStep sg2 sg1)} holds.
Consider a node \ls{p} whose $\alpha$-potential has increased during
the step, \textit{i.e.}, \lstinline{($\alpha$_pot sg1 p) $<_{\tt P}$ ($\alpha$_pot sg2 p)}.  This means, by definition of
\lstinline{$\alpha$_pot}, that \ls{p} is disabled in \ls{sg1} (its
potential is 0) and becomes enabled in \ls{sg2} (its potential
becomes %
\lstinline{(depth sg2 p)$>0$}).

To show the local criterion, we exhibit a down-path in the tree $T$ from
\ls{p} to some leaf, which contains a witness node that is
$\alpha$-enabled in \ls{sg1} and $\alpha$-disabled in next
configuration \ls{sg2}.
We prove the result in two steps. First, we exhibit a child of node
\ls{p}, \ls{child}, which necessarily executes the first action of its
algorithm during the step.  This is proven by induction on the
neighbors of \ls{p} using the fact that \ls{run} only depends on the
states of the children of \ls{p} in the tree $T$.  Next, we prove the
following lemma:
\begin{lstlisting}
  Lemma moving_node_has_disabled_desc: $\forall$(child: Node), 
    alpha ((getEnv sg1) child) $\neq$ alpha ((getEnv sg2) child) $\to$
    $\exists$(desc: Node), 
      ($\exists$(path: list Node), directed_tree_path child path desc) $\wedge$ 
      $\alpha$_enabled sg1 desc $\wedge$ $\lnot$ $\alpha$_enabled sg2 desc
\end{lstlisting}
where \ls{directed_tree_path} checks that \ls{path} is actually a path
from \ls{child} to \ls{desc} in the directed spanning tree $T$.  The
lemma states that when the node \ls{child} $\alpha$-moves, it is
down-linked in $T$ to a node which was $\alpha$-enabled and becomes
$\alpha$-disabled, during the step. Hence, the lemma provides the
witness node required to prove the local criterion.

The lemma is proven by induction on \ls{(WF_child child)},
\textit{i.e.}, on the down-paths from \ls{child} in $T$.  Consider a
node in such a path which is enabled in \ls{sg1} and that $\alpha$-moves
during the step from \ls{sg1} to \ls{sg2}. We have two cases.
\begin{itemize}
\item Either it becomes disabled in \ls{sg2}: this is the base case
  of the induction, taking \ls{desc} as \ls{child} and \ls{path} empty.
\item Or it is still enabled in \ls{sg2}: for this case, we prove
  that any node that executed the first action of the algorithm in
  \ls{sg1} but is still $\alpha$-enabled in \ls{sg2} has a child in
  $T$ which has also $\alpha$-moved (the proof is based on induction
  on the children of the node). This result provides the induction
  step of the proof.
\end{itemize}

\subsubsection{Global Criterion for $\alpha$\lstinline{_SafeStep}s} 

The global criterion requires to find a witness node whose
$\alpha$-potential differs between \ls{sg1} and \ls{sg2}. We show that
there exists a node \ls{p} with $\alpha$-potential \ls{(depth sg1 p)}
in \ls{sg1} (such a potential is necessarily greater than 0), and
$\alpha$-potential 0 in \ls{sg2}. Namely, \ls{p} is $\alpha$-enabled
in \ls{sg1}, but $\alpha$-disabled in \ls{sg2}. The proof uses the
fact that at least one node, say \ls{q}, has $\alpha$-moved during the
step (see definition of \lstinline{QTrans$_{\alpha}$}).  Then, we use
Lemma \ls{moving_node_has_disabled_desc} again to exhibit a witness
node \ls{p} (on a given down-path of $T$ from \ls{q}) which is
$\alpha$-enabled in \ls{sg1}, but $\alpha$-disabled in \ls{sg2}.

\subsubsection{Conclusion for $\alpha$\lstinline{_SafeStep}s}

Local and global criteria being proven, we directly obtain Theorem
\lstinline{$\alpha$_safe_inclusion} from
Section~\ref{sec:termination-theorem}. For safe steps involving the second
action (predicate $\parCl$\lstinline{_SafeStep}) and the third action (predicate $\head$\lstinline{_SafeStep}),
we use exactly the same method, see in the next sections.


\subsection{$\parCl$\lstinline{_SafeStep}}

For the second action, we build safe steps,
called \lstinline{$\parCl$_SafeStep}, by instantiating the predicate
\lstinline{QTrans} as follows: for any two safe configurations
\ls{sg}
and \ls{sg'},
\begin{lstlisting}
  QTrans$_{\parCl}$ sg' sg := $\lnot$QTrans$_{\alpha}$ sg' sg $\wedge$ 
    $\exists$(p: Node), $\parCl$_enabled sg p $\wedge$ has_moved sg' sg p 
\end{lstlisting}
where \lstinline{$\parCl$_enabled} is exactly the guard of the second
action. As before, \lstinline{QTrans$_{\parCl}$} is proven decidable
and when %
\lstinline{($\parCl$_enabled sg p)}, we say that node \ls{p} is
{\em $\parCl$-enabled in the safe configuration \ls{sg}}.
In a \lstinline{$\parCl$_SafeStep}, we require that no node executes its first action
(nodes can be $\alpha$-enabled, but do not move) and at least a node
executes its second action (other nodes can execute or not
their third action).
We then show the theorem
\begin{lstlisting}
  Theorem $\parCl$_safe_inclusion: $\forall$(sg1 sg2: safeEnv), 
    $\parCl$_SafeStep sg2 sg1 $\to$ (Pot$_{\parCl}$ sg2) $\ltM$ (Pot$_{\parCl}$ sg1)
\end{lstlisting}
using exactly the same method as for $\alpha$.

During any \lstinline{$\parCl$_SafeStep} between safe configurations \ls{sg1} and
\ls{sg2}, no node executes its first action, as expressed in
\lstinline{QTrans$_{\parCl}$}. Hence, all the values of $\alpha$ stay
unchanged. As a consequence, for every node \ls{p}, the macro
$\ParCLR($\ls{p}$)$ from Algorithm~\ref{alg:clk} (which only depends on the
values of $\alpha$) has the same result when evaluated at both safe
configurations \ls{sg1} and \ls{sg2}. This ensures that
\begin{enumerate}
\item \label{it:a} a node which executes its second action during the
  \lstinline{$\parCl$_SafeStep} from \ls{sg1} to \ls{sg2} is no longer
  $\parCl$-enabled in \ls{sg2};
\item \label{it:b} a $\parCl$-disabled node in \ls{sg1} remains
  $\parCl$-disabled in \ls{sg2}.
\end{enumerate}
From those observations, we define the $\parCl$-potential of a node
\ls{p} in safe configuration \ls{sg}, \lstinline{($\parCl$_pot sg p)}, as 1 if \ls{p} is $\parCl$-enabled in \ls{sg} and 0,
otherwise. Now, criteria from Section~\ref{sec:termination-theorem}
are straightforward. Indeed, we have
\begin{itemize}
\item the \textit{local criterion}, since the $\parCl$-potential of a
  node cannot increase (\textit{i.e.} switch from 0 to 1, see
  (\ref{it:b}) above);
\item the \textit{global criterion}, since, due to %
  \lstinline{QTrans$_{\parCl}$}, there exists a node which executes
  its second action: from (\ref{it:a}) above, its $\parCl$-potential
  differs between \ls{sg1} and \ls{sg2} since it switches from 1 to 0.
\end{itemize}


\subsection{$\head$\lstinline{_SafeStep}}
\label{sec:hd-term}

For the third action, we build safe steps, called
\lstinline{$\head$_SafeStep}, by instantiating the predicate
\lstinline{QTrans} as follows: for any two safe configurations \ls{sg}
and \ls{sg'},
\begin{lstlisting}
  QTrans$_{\head}$ sg' sg := $\lnot$QTrans$_{\alpha}$ sg' sg $\wedge$ $\lnot$QTrans$_{\parCl}$
\end{lstlisting}
As before, \lstinline{QTrans$_{\head}$} is proven decidable. We also
denote by \lstinline{$\head$_enabled} the guard of the third action and
when %
\lstinline{($\head$_enabled sg p)}, we say that node \ls{p} is
{\em $\head$-enabled in the safe configuration \ls{sg}}.
In a \lstinline{$\head$_SafeStep} between safe configurations, we require that no node
executes its first or second action (nodes can be $\alpha$-enabled or
$\parCl$-enabled, but do not move in that case); furthermore, as
enforced by the predicate
\ls{Step}, at least one node executes: therefore it executes its
third action.
We then show the theorem:
\begin{lstlisting}
  Theorem $\head$_safe_inclusion: $\forall$(sg1 sg2: safeEnv), 
    $\head$_SafeStep sg2 sg1 $\to$ (Pot$_{\head}$ sg2) $\ltM$ (Pot$_{\head}$ sg1)
\end{lstlisting}
using exactly the same method as for $\alpha$.

\subsubsection{$\parCl$-path}

The fact that third actions terminate is due to the fact that $\head$
values are computed along the paths made of $\parCl$ pointers. We call
those paths {\em $\parCl$-paths} and we define them using the relation
\ls{pcl_rel sg}, where \ls{sg} is a safe configuration: two nodes
\ls{p} and \ls{q} are related \textit{via} \ls{(pcl_rel sg)} \ls{(pcl_rel
  sg q p)} when \ls{p} is $\alpha$-disabled and $\parCl$-disabled in
\ls{sg}, and when its $\parCl$-pointer points to \ls{q} in
\ls{sg}. A $\parCl$-path between two nodes \ls{P} and \ls{Q}, if 
exists, is then the list of nodes, from \ls{P} to \ls{Q}, built using
the transitive closure of \ls{(pcl_rel sg)}.

We show that the relation \ls{(pcl_rel sg)} is well-founded, for any
safe configuration \ls{sg}.  First, we observe from the algorithm that
when two nodes \ls{p} and \ls{q} are related, {\em i.e.} \ls{(pcl_rel sg q p)},
either \ls{p} is short and \ls{q} is its parent in $T$, or \ls{p} is
tall and \ls{q} is a child of \ls{p} in $T$.  Then, we split the proof
into two parts.
\textit{For tall nodes,} we prove that for any two related nodes
\ls{p} and \ls{q}, such that \ls{(pcl_rel sg q p)},
  we have: \ls{q} is tall whenever \ls{p} is.
Hence, for a given node \ls{p}, we can prove \ls{(Acc
  (pcl_rel sg) p)} directly from induction on \ls{(WF_child p)}.
\textit{For short nodes,} a short node can be linked using
\ls{(pcl_rel sg)} to a short (like $m$ in Configuration (I) of Figure
\ref{fig:clr}) or a tall node (like $i$ in Configuration (I) of Figure
\ref{fig:clr}). To prove \ls{(Acc (pcl_rel sg) p)} for a given node
\ls{p}, we proceed again by induction, following the $\parCl$-path of
\ls{p}, on \ls{(WF_par p)}. But it may occur that the path reaches a
tall node, in which case, we use the previous result for tall nodes to
be able to conclude the induction case.

From the well-foundedness of \ls{(pcl_rel sg)} for any safe
configuration \ls{sg}, we can inductively define %
\lstinline{(dist_hd sg p)} as the length of the $\parCl$-path of
\ls{p} in \ls{sg}.  Since $\parCl$-pointers are constant in any
\lstinline{$\head$_SafeStep}, \lstinline{dist_hd} has the same value in any two safe
configurations \ls{sg1} and \ls{sg2} as far as they are linked by a
\lstinline{$\head$_SafeStep}, \textit{i.e.,} \lstinline{($\head$_SafeStep} \ls{sg2} \ls{sg1)} holds.
From this result, we can show that
\begin{enumerate}
\item \label{it:hda} If a node \ls{p} is $\head$-enabled in \ls{sg1},
  executes, and remains $\head$-enabled in \ls{sg2}, then this means
that \ls{p} has a successor in its $\parCl$-path and this successor is
also $\head$-enabled in \ls{sg1} and executes during the step.
\item \label{it:hdb} As the $\parCl$-path is finite, this proves
  (by a structural induction on \ls{(Acc (pcl_rel} \ls{sg1)} \ls{p)})
  that \ls{p} is 
  necessarily linked in its $\parCl$-path to a node \ls{q}, which is
  $\head$-enabled in \ls{sg1}, executes, and becomes $\head$-disabled
  in \ls{sg2}; furthermore we have that
    \begin{lstlisting}
      dist_hd sg1 q < dist_hd sg1 p
    \end{lstlisting}
\item \label{it:hdc} Therefore, from (\ref{it:hda}) and
  (\ref{it:hdb}), a node \ls{P}, which is $\head$-enabled in \ls{sg1}
  and executes during the step, is necessarily linked \textit{via} its
  $\parCl$-path to a node \ls{Q} such that \ls{Q} is $\head$-enabled in
  \ls{sg1}, $\head$-disabled in \ls{sg2}, and %
    \begin{lstlisting}
      dist_hd sg1 Q $\leq$ dist_hd sg1 P
    \end{lstlisting}
    This node can be \ls{P} itself if \ls{P} becomes $\head$-disabled in
    \ls{sg2}, or a node which is further along in the $\parCl$-path.
\end{enumerate}

\subsubsection{$\head$-Potential}

We could have used \ls{dist_hd} to build the $\head$-potential but
results from Section~\ref{sec:proof-convergence-part-general} assume
decreasing potential, whereas this one would have been
increasing. Instead, we prove the existence of a natural number
\ls{NN} such that proposition
\begin{lstlisting}
  HNN := $\forall$(sg: safeEnv) (p: Node), NN > dist_hd sg p
\end{lstlisting}
is true. We use the tools about quantitative properties (see
Section~\ref{sec:proof-counting-part}) to achieve the proof. We set
\ls{NN} as $(n + 1)$, where $n$ is the number of nodes in the network,
using the list \ls{all_nodes} to show the existence of $n$. We
prove \ls{HNN} using the fact that in a $\parCl$-path, each node
occurs at most once; this comes from the well-foundedness of
\ls{(pcl_rel sg)} from which we can infer that a $\parCl$-path
contains no loop.

Finally, for a given safe configuration \ls{sg} and a given node
\ls{p}, we pick its $\head$-potential to be \ls{(NN - dist_hd sg p)}
if \ls{p} is $\head$-enabled in \ls{sg} and 0 otherwise.

\subsubsection{Global Criterion for $\head$\lstinline{_SafeStep}s}

The global criterion requires to exhibit a node whose potential has
changed from \ls{sg1} to \ls{sg2}: we look for a node which is
$\head$-enabled in \ls{sg1} (potential is $>0$) and $\head$-disabled
in \ls{sg2} (potential is 0). From \lstinline{Step}, there exists a
node \ls{p} which executes during the step and
\lstinline{QTrans$_{\head}$} guarantees that it uses its third action.
We directly use the result~(\ref{it:hdc}) above: there exists a node
\ls{q} in the $\parCl$-path of \ls{p} which is $\head$-enabled in
\ls{sg1} and $\head$-disabled in \ls{sg2}, hence its $\head$-potential
changes during the step.

\subsubsection{Local Criterion for $\head$\lstinline{_SafeStep}s}

To show the local criterion, we consider a node \ls{p} whose
$\head$-potential increases during the step. Specifically, a node which is
$\head$-disabled in \ls{sg1} and $\head$-enabled in \ls{sg2}. We prove
that this situation is possible only if \ls{p} has a successor \ls{p'}
in its $\parCl$-path such that \ls{p'} $\head$-executes during the
step. Note that \ls{(dist_hd sg1 p)} is greater than \ls{(dist_hd sg1
  p')}. We use the result~(\ref{it:hdc}) above, which ensures the
existence of a node \ls{q} such that
\begin{itemize}
\item \ls{q} is $\head$-enabled in \ls{sg1} and $\head$-disabled in
  \ls{sg2} (hence its $\head$-potential changes during the step), and
\item %
  \lstinline{(dist_hd sg1 q $\leq$ dist_hd sg1 p' = (dist_hd sg1 p) - 1)}.
  Hence, \\ \lstinline{(pot_hd sg2 p = NN - dist_hd sg1 p <  pot_hd sg1 q = NN - dist_hd sg1 q)}.
\end{itemize}


\subsection{Termination}

We proved the safe inclusions for the three kinds of safe steps, so we
can apply the lexicographical order method with three dimensions.  It
requires to show that assumptions about priorities, as encoded by the
lexicographical order, conform to the algorithm and to the definitions
of $\alpha$-steps, $\parCl$-steps, and $\head$-steps; this is
translated into the assumption \ls{Hdisjoint} instantiated at two
levels, namely, we need to verify that
\begin{lstlisting}
  Hdisjoint$_{cl}$ := $\forall$(sg' sg: safeEnv),
    $\parCl$_SafeStep sg' sg $\vee$ $\head$_SafeStep sg' sg $\to$ 
    (Pot$_{\alpha}$ sg') $\eqM$ (Pot$_{\alpha}$ sg)
    $\bigwedge$
    $\head$_SafeStep sg'sg $\to$ (Pot$_{\parCl}$ sg') $\eqM$ (Pot$_{\parCl}$ sg)
\end{lstlisting}
(when a $\parCl$-step or a $\head$-step occurs, no $\alpha$-potential
changes and when a $\head$-step occurs, no $\parCl$-potential
changes); the validity of the condition comes directly from the values
of \lstinline{QTrans$_{\parCl}$} and \lstinline{QTrans$_{\head}$}.

From \lstinline{Hdisjoint$_{cl}$},
\lstinline{$\alpha$_safe_inclusion}, \lstinline{$\parCl$_safe_inclusion}
and \lstinline{$\head$_safe_inclusion}, we apply Lemma
\ls{union_lex_wf3} and obtain that
\begin{lstlisting}
  well_founded ($\alpha$_SafeStep $\cup$ $\parCl$_SafeStep $\cup$ $\head$_SafeStep).
\end{lstlisting}
Finally, we proved the equivalence between the relations
\lstinline{SafeStep} and %
\lstinline{($\alpha$_SafeStep} \lstinline{$\cup$}
\lstinline{$\parCl$_SafeStep} \lstinline{$\cup$}
\lstinline{$\head$_SafeStep)}: this directly comes from the values of
the predicates \lstinline{QTrans$_{\alpha}$},
\lstinline{QTrans$_{\parCl}$} and \lstinline{QTrans$_{\head}$}.  This
ends the proof and concludes that %
\lstinline{(well_founded} \ls{SafeStep)}. Using Lemma
\ls{Acc_Algo_Multiset}, we obtain the desired property:
\begin{lstlisting}
  Theorem $\algocl{k}$_termination: $\forall$(g: Env), Assume$_{cl}$ g $\to$ Acc Step g.
\end{lstlisting}

\section{Partial Correctness of \texorpdfstring{$\algocl{k}$}{C(k)}}
\label{sec:corr-algocl}

We develop a certified proof of the \ls{P_correctness} property of
$\algocl{k}$. Namely, we show the partial correctness of $\algocl{k}$,
as far as \lstinline{Assume$_{cl}$} is satisfied:
\begin{lstlisting}
Theorem $\algocl{k}$_at_terminal: $\forall$(g: Env), Assume$_{cl}$ g $\to$ terminal g $\to$ $\dsP_{cl}$ g.
\end{lstlisting}
This goal is divided into three subgoals.  First, we prove the partial
correctness of the first actions; this is achieved by proving that
once the algorithm has converged, the $\alpha$-values allow to define
a $k$-hop dominating set. Second, we prove, that after termination,
the strong $k$-clustering specification holds (\textit{i.e.}, each
node knows the identifier of its clusterhead and the channel
corresponding to its parent link in its $k$-cluster tree).
Third, we show that any terminal configuration contains at most
$\lfloor \frac{n-1}{k+1} \rfloor + 1$ $k$-clusters.

\subsection{Proof for a $k$-hop Dominating Set}
\label{sec:proof-k-dominating}

\subsubsection{Values of \texorpdfstring{$\alpha$}{alpha} are in range \texorpdfstring{$\{0, ..., 2k\}$}{\{0..2k\}}}
As a preliminary result, the value of $\alpha$ at a node \ls{p} is in
range $\{0, ..., 2k\}$ after \ls{p} participates in any step and also
when the system is in a terminal configuration.  The proof shows that
the value returned by macro $Alpha(\texttt{p})$ is in range $\{0,...,
2k\}$: this is proven using a case analysis on $\MaxAShort(\texttt{p})
+ \MinATall(\texttt{p}) > 2 k - 2$ and the fact that, by definition,
$-1 \le \MaxAShort(\texttt{p}) \le k-1$ and $k \le
\MinATall(\texttt{p}) \le 2k+1$.

\subsubsection{Proof for $\clrdom$}
\label{sec:proof-dom}

We prove that the set $\clrdom$, made of all the nodes $p$ such that
$\kDominator(p)$ holds, is a $k$-hop dominating set in any terminal
configuration, namely we need to check the existence of a path in $G$
between any node \ls{p} and any node \ls{kdom} of $\clrdom$, such that
this path is of length at most $k$.
To be usable by the rest of the proof, we show a bit more. Actually,
once the $\alpha$-values allow to define a $k$-hop dominating set,
they also exhibit routing paths between each node and one of its
witnesses in $\clrdom$ by choosing one of the possible path: this
choice is made using the ordering on channels and the $\min_{<_C}$
operator. Therefore, in this part, we prove that \emph{any} such
possible routing path has length at most $k$.

\paragraph{Tree Paths}

To achieve this property, the algorithm builds tree paths of
particular shape: those paths use edges of $T$ in both
directions. Precisely, these edges are defined using relation
\ls{(is_kDom_edge g)}, in a given configuration \ls{g}, which depends
on $\alpha$-values: for any short node \ls{s}, we select the edge from
\ls{p} to \ls{s}, where \ls{p} is the parent of \ls{s} in $T$
(\ls{Par}), {\em i.e.}, \ls{(is_kDom_edge g p s)} holds; while for any
tall node \ls{t} which is not in $\clrdom$, we select every edge from
\ls{c} to \ls{t}, where \ls{c} is a child of \ls{t} in $T$ such that 
\lstinline{($\alpha$ (g c) = $\alpha$ (g t) -1)}, {\em i.e.},
\ls{(is_kDom_edge g c t)} holds. The relation \ls{(is_kDom_edge g)}
defines a subgraph of $G$ called the \ls{kdom}-graph of
\ls{g}. (Remark that all directed edges in graphs of
Figure~\ref{fig:clr} appear in their associated \ls{kdom}-graph, yet
in opposite sense.)

The rest of the analysis is conducted assuming a terminal
configuration \ls{(g: Env)} which contains a rooted spanning tree
built upon a bidirectional graph, namely such that %
\lstinline{(Assume$_{cl}$ g)} and \ls{(terminal g)} hold.  We aim at
proving the following result:
\begin{lstlisting}
  OK_dom g p :=
    ($\exists$(kdom: Node), ($\kDominator$ g kdom) $\wedge$ 
      $\exists$(path: list Node), is_kDom_path g path kdom p)
    $\wedge$ 
    $\forall$(kdom: Node) (path: list Node), 
      is_kDom_path g path kdom p $\to$ (length path) $\leq$ $k$.

  Theorem kDom_correctness: $\forall$(p: Node), OK_dom g p.
\end{lstlisting}
where \ls{is_kDom_path} checks that its parameter \ls{path} is a path
in the \ls{kdom}-graph of configuration
\ls{g} between \ls{kdom} and \ls{p}.
The proof of \ls{(OK_dom g p)} for any node \ls{p} is split into two
cases, depending on whether \ls{p} is tall or short. Actually, we
prove by straightforward induction on \ls{i} that, for any node \ls{p}
and any natural \ls{i}, such that %
\lstinline{($\alpha$ (g p) = $k$ + i)} %
(resp. \lstinline{($\alpha$ (g p) = $k$ - i)}), %
the property \ls{(OK_dom g p)} holds -- where the length of \ls{path}
is at most \ls{i}.

\paragraph{Proof for Tall Nodes}

For case \ls{(i = 0)}, \ls{p} satisfies \lstinline{$\kDominator$(p)}
and any path from \ls{p} to \ls{p} in the \ls{kdom}-graph of \ls{g} has 
length 0.  For case \ls{(i = j + 1)}, as %
\lstinline{($\alpha$ (g p) = $k$ + i)} is positive, we can prove using
a case analysis on $\MaxAShort(\mathtt{p}) + \MinATall(\mathtt{p}) \le
2 k - 2$, that there exists a child \ls{q} of \ls{p} with %
\lstinline{($\alpha$ (g q) = $k$ + j)} on which we can apply the
induction hypothesis. This exhibits a path in the \ls{kdom}-graph of
\ls{g} from some $k$-hop dominator \ls{kdom} to \ls{q}. Since \ls{p} is
the parent of \ls{q} in $T$ ({\em i.e.}, \ls{q} is a child of \ls{p}), we obtain a path from \ls{kdom} to \ls{p}
at \ls{g}.

Looking at the second part of the result, we have to prove that any
\ls{path} in the \ls{kdom}-graph of \ls{g} has length at most \ls{i}.
Then, either \ls{path} is empty and its length is 0, or we can
decompose it into a \ls{kdom}-path \ls{path'} followed by some node
\ls{q} and then \ls{p}, such that there is a \ls{kdom}-edge between
\ls{q} and \ls{p}. Using the definition of \ls{kdom_edge}, we obtain
that %
\lstinline{($\alpha$ (g q) = $\alpha$ (g p) - 1 = $k$ + j)}: hence we
apply again the induction hypothesis to \ls{q} and obtain that
\ls{path'} has length at most \ls{j}; hence \ls{path} has length at
most \ls{(j + 1)}.

\paragraph{Proof for Short Nodes}

The case \ls{(i = 0)} is already proven by the above result for tall
nodes.  We now look at case \ls{(i = j + 1)}. When \ls{p} is the root
of $T$, then \lstinline{$\kDominator$(p)} holds and any path in the
\ls{kdom}-graph of \ls{g} whose terminal extremity is $p$ is empty. We
now assume that \ls{p} is non-root.

For the first part of the property (looking for a witness
\lstinline{kdom${}\in\clrdom$} and a path from \ls{kdom} to \ls{p}), we pick
the parent \ls{q} of \ls{p} in $T$. We can show that %
\lstinline{($\alpha$ (g p) $\leq$ $\alpha$ (g q) + 1)} and that
\ls{(is_kDom_edge g q p)}. If \ls{q} is also short, the induction
hypothesis applies directly, otherwise (\ls{q} is tall), the above
property \ls{(OK_dom g q)} holds. In both cases, this provides a
witness node \ls{kdom} in $\clrdom$ and a path from \ls{kdom} to
\ls{q} in the \ls{kdom}-graph of \ls{g}; we add to \ls{path} the
\ls{kdom}-edge from \ls{q} to \ls{p} to build a path from \ls{kdom} to
\ls{p} in \ls{kdom}-graph.

For the second part of the property, we consider a \ls{kdom}-path
\ls{path} from some node to \ls{p}. Either this path is empty, in
which case, the property trivially holds, or we can decompose it into
a sub-path \ls{path'} and an edge from some node \ls{q} to \ls{p} in
\ls{kdom}-graph. From the definition of \ls{kdom_edge},
\ls{q} is the parent of \ls{p} in $T$. If \ls{q} is short, we apply
the induction hypothesis to obtain that the length of \ls{path'} is at
most \ls{j}. Otherwise \ls{q} is tall and we have two cases:
\begin{itemize}
\item If \lstinline{$\MaxAShort$(q) + $\MinATall$(q) > 2 $k$ - 2},
  then we have %
  \begin{lstlisting}
  ($\alpha$ (g q) = $\MaxAShort$(q) + 1 $\le$ $k$).
  \end{lstlisting}
  The fact that \ls{q} is tall implies %
  \lstinline{($\alpha$ (g q) = $k$)}. Hence, the \ls{path'} has length
  $1$.

\item Otherwise, since \ls{q} is tall, from the result \ls{(OK_dom g
    q)} above, \ls{path'} has length at most %
  \lstinline{($\alpha$ (g q) - $k$)}.
  Now, since \ls{p} is a short child of \ls{q} in $T$, we
  have that \begin{lstlisting}
    ($\alpha$ (g p) $\le$ $\MaxAShort$(q)).
  \end{lstlisting}
  We also
  have \lstinline{($\alpha$ (g q) = $\MinATall$(q) + 1)}.  Combining
  all, we obtain that %
  \begin{lstlisting}
    ($\alpha$ (g q) - $k$ $\leq$ $k$ - $\MaxAShort$(q) - 1 $\leq$ j).
  \end{lstlisting}
  Hence, \ls{path'} has a length at most \lstinline{j} and \ls{path}
  at most \ls{(j + 1)}.
\end{itemize}



\subsection{Proof for $k$-Clustering}
\label{sec:proof-k-clustering}

We explain here the proofs of the parts \ls{kCluster_strong} and \ls{partition_OK}
of the specification $\dsP_{cl}$. We instantiate
\begin{itemize}
\item \ls{clusterHeadID} as $\head$,
\item \ls{clusterParent} is set to $\parCl$ and
\item \ls{(kCluster g h p)} is defined by %
  \lstinline{(eqId $\id$(h) (clusterHeadID (g p)))} for any
  configuration \ls{(g: Env)} and nodes \ls{h p: Node)}.
\end{itemize}
For the rest of the analysis, we fix a configuration \ls{(g: Env)}
such that \lstinline{(Assume$_{cl}$ g)} and \ls{(terminal g)} hold.
As a preliminary remark, we prove that every node in a cluster path
declares the same clusterhead in \ls{g}:
\begin{lstlisting}
  Lemma same_hd: $\forall$(b e: Node) (path: list Node), 
    cluster_path g b path e $\to$ eqId ($\head$ (g b)) ($\head$ (g e)).
\end{lstlisting}
The proof is an easy induction on the list \ls{path} using the
expression of macro $\ParCLR$.  This lemma ensures that the predicates
\ls{(cluster_path g)} and \ls{(agreed_cluster_path g)} are equivalent
(see Section~\ref{sec:algo-spec}).


\subsubsection{Relation \ls{is_cluster_parent}}

We first study the relation \ls{(is_cluster_parent g)}. We show that
it is included into the \ls{kdom}-edges of the \ls{kdom}-graph:
\begin{lstlisting}
  inclusion (is_cluster_parent g) (is_kDom_edge g)
\end{lstlisting}
The proof is a (quite long) case analysis based on the fact that
\ls{g} is terminal and on the macros $\Alpha$ and $\ParCLR$.
We also show that this relation is well-founded:
\begin{lstlisting}
  well_founded (is_cluster_parent g)
\end{lstlisting}
Actually, we proved it during the termination proof (see the part about
$\head$\lstinline{_SafeStep}s in
Section~\ref{sec:hd-term}). Precisely, we showed that \ls{(pcl_rel
  g')} is well-founded for any configuration \ls{g'}. Now, since
\ls{g} is terminal, \ls{(is_cluster_parent g)} is included in
\ls{(pcl_rel g)}, and we obtain the well-foundedness.

\subsubsection{$\kDominator$ and \ls{clusterHead}}

We manage to prove the equivalence between %
\lstinline{($\textit{kDom}$}- \lstinline{$\textit{inator}$ g)} and \ls{(clusterHead g)}:
\begin{lstlisting}
  $\forall$(p: Node), $\kDominator$ g p $\leftrightarrow$ clusterHead g p.
\end{lstlisting}
First, we transform this goal into
\begin{lstlisting}
  $\forall$(p: Node), eqoptionA eqC ($\parCl$ (g p)) $\perp$ $\leftrightarrow$ clusterHead g p.
\end{lstlisting}
since we can prove that %
\lstinline{($\kDominator$ g p $\leftrightarrow$ eqoptionA eqC ($\parCl$ (g p)) $\perp$)}. 
Indeed, this latter result is based on the fact the
\lstinline{($\parCl$ (g p))} is equal to \lstinline{$\ParCLR $(p)} in
the terminal configuration \ls{g}; the proof uses a case analysis
which treats the case when \ls{p} is short easily. For the case when
\ls{p} is a non-root tall node, it requires to prove that
$\MinIDMinATall$ actually returns a tall child of \ls{p} which is a
quite tricky intermediate result, based on the definition of
$\MinIDMinATall$.

Back to the goal above, the direct part of the equivalence comes
directly from the fact that \lstinline{($\parCl$ (g p))} is $\perp$,
using the expression of $\ParCLR$ and the fact that \ls{g} is
terminal. Now, we focus on the reverse part of the equivalence and we
assume that \lstinline{$\id$(p)} and \lstinline{($\head$ (g p))} are
equal.

As \ls{(is_cluster_parent g)} is a well-founded and decidable
relation, and as there exists a finite number of nodes in the network,
for any node \ls{p}, we can build the maximal cluster path called
\ls{(path: list Node)} from \ls{p}: it reaches some node called
\ls{(h: Node)} which has no cluster parent pointer; hence \ls{path}
and \ls{h} satisfy:
\begin{lstlisting}
        cluster_path h path p $\wedge$ $\forall$(x: Node), $\lnot$cluster_parent x h.
\end{lstlisting}

When \ls{path} is empty, proof is done, since \ls{p} and \ls{h} are
the same node with no cluster parent pointer. Otherwise \ls{path} is
not empty: we show that this case is not possible since it yields a
contradiction. Indeed, nodes in a non-empty cluster path are all
different (we prove that \ls{path} contains no loop since
\ls{(is_cluster_parent g)} is a well-founded relation over a finite
set), hence have all different identifiers (since \ls{uniqueId} is
assumed).  This ensures that \lstinline{$\id$(h)} and
\lstinline{$\id$(p)} are different.
From Lemma \ls{same_hd}, we obtain that \lstinline{($\head$ (g h))}
and \lstinline{($\head$ (g p))} are equal and since %
\lstinline{($\parCl$ (g h))} is $\perp$, the expression of $\ParCLR$
ensures that \lstinline{($\head$ (g h))} is \lstinline{$\id$(h)}.
Those three equations yield a contradiction with the fact that
\lstinline{$\id$(p)} and \lstinline{($\head$ (g p))} are equal, as
assumed at the beginning of the proof.

\subsubsection{Proof of \ls{kCluster_strong}}
The first line of \ls{kCluster_strong} (see
page~\pageref{kcluster_strong}, for the definition) is exactly given by
the instantiation of \ls{kCluster}. For the second line, assuming
\ls{(kCluster g b e)}, for any two nodes \ls{b} and \ls{e}, we define
a function that builds the maximal cluster path \ls{path} from \ls{e}
to \ls{b} and we prove that \ls{b} is a clusterhead in \ls{g}:
\ls{(clusterHead g b)}. As \ls{(cluster_path g)} is included into
\ls{(kdom_path g)}, we are able to use the theorem which asserts that
\ls{(OK_dom g e)} and prove that any cluster path has length at most
$k$.

\subsubsection{Proof of \ls{partition_OK}}
(see page~\pageref{partition_ok} for the definition)
Let \ls{p} be a node. As before, we can build the maximal cluster
path, called \ls{path} from \ls{p} to its clusterhead \ls{h}. From
Lemma \ls{same_hd}, \ls{p} declares \ls{h} as its clusterhead and so
does \ls{h}. Hence, \ls{(kCluster g h p)} holds.  Now to prove
uniqueness of clusterheads, we assume \ls{(kCluster g h p)} and
\ls{(kCluster g h' p)} and the goal is to prove that nodes \ls{h} and
\ls{h'} are equal. This goal is transformed into %
\lstinline{(eqId $\id$(h) $\id$(h'))} using the uniqueness assumption
\ls{uniqueId}. As the assumptions on \ls{kCluster} both expands into
\lstinline{(eqId $\id$(b) ($\head$ (g n)))} and %
\lstinline{(eqId $\id$(b')($\head$ (g n)))} we are done by
transitivity.






\subsection{Proof for counting}
\label{sec:proof-counting}

In this section, we formally prove the property \ls{count_OK} which
states that $(n-1) \ge (k+1)(|\texttt{CH}|-1)$ where \ls{CH} has been
defined as the set of clusterheads. Intuitively, this means that all
but one element of \ls{CH} have been chosen as clusterheads by at least
$k+1$ distinct nodes each. Actually, we prove that $\clrdom$ has this
property: $(n-1) \ge (k+1)(|\clrdom|-1)$ and then use the equivalence
between \ls{clusterHead} and \lstinline{$\kDominator$} to conclude.

The proof outline is the following.  First, we assume a terminal
configuration \ls{(g: Env)}, \ls{(terminal g)}, such that
\lstinline{(Assume$_{cl}$ g)} holds. The existence of the natural
number $n$ (number of nodes) is given using the results in
Section~\ref{sec:numb-elem-lists} about the number of elements in the
list \ls{all_nodes}. Similarly, the existence of the natural number
$|\clrdom|$ (number of nodes in $\clrdom$) is given using the above
results applied to list \ls{all_nodes} and predicate function
\lstinline{(fun p: Node => ($\kDominator$ (g p) = true))}.

We define as \emph{regular head} each node \ls{h} such that $\alpha$
equals $k$ in \ls{g}
\begin{lstlisting}
  RegHead h := ($\alpha$ (g h) = $k$)
\end{lstlisting}
and the set of regular
heads as %
\lstinline!RegHeads := { h: Node | RegHead h }!. Note that by
definition, \ls{RegHeads} is included in $\clrdom$. Again, we prove
the existence of the natural number $rh$ which represents the number
of nodes in \ls{RegHeads} using list \ls{all_nodes} and predicate
\ls{RegHead}.

We also define a \emph{regular node} as a node which declares a
regular head as clusterhead.  In practice, regular nodes are tall or
have a tall ancestor in $T$. In case the root is a short clusterhead,
they cannot be in its cluster.  We define predicate
\ls{HasTallAncestor} as an inductive predicate which selects any node
which has a tall ancestor in $T$ (\textit{i.e.} such that there is a
directed path in $T$ from \ls{p} to the root $\r$ that contains a node
with $\alpha$ at least $k$). The set of regular nodes is defined by:
\ls{RegNodes :=} \ls{\{ p: Node | HasTallAncestor p \}}.  Again, we
prove the existence of the natural number $rn$ which is the number of
nodes in \ls{RegNodes}.
Now, we prove the following theorem:
\begin{lstlisting}
  Theorem simple_counting: $rn \geq (k + 1) rh$.
\end{lstlisting}
Using results from the library on cardinality of sets and lists, this
theorem is reduced to
\begin{lstlisting}
  Card Smaller ($\mathcal{M}_{k+1}$ $\times$ RegHeads) RegNodes.
\end{lstlisting}
This latter proposition is proven by constructing a relation
\ls{Rcount} from pairs of natural numbers $i\in\{0, ..., k\}$ and
\emph{regular heads} to regular nodes, such that: for a regular head
\ls{h}, some $i\in\{0, ..., k\}$ and a regular node %
\lstinline{p$_i$}, %
\lstinline{(Rcount ($i$, h) p$_i$)} holds if and only if %
\lstinline{($\alpha$ (g p$_i$) = $i$)} and %
\lstinline{p$_i$} designates \ls{h} as clusterhead (\textit{i.e.},
there is a maximal path from \lstinline{p$_i$} to \ls{h} in the
\ls{kdom}-graph of \ls{g}). We show that \ls{Rcount} is actually an
injection of domain %
\lstinline{($\mathcal{M}_{k+1}$ $\times$ RegHeads)}. Indeed, for any
pair \lstinline{($i$, h)}, there is a node \lstinline{p$_i$} such
that %
\lstinline{($\alpha$ (g p$_i$) = $i$)} which designates \ls{h} as
clusterhead; the proof is carried out by induction on values of $i$.
Intuitively, this implies that there is a path of length $k+1$ in
the \ls{kdom}-graph of \ls{g} linking \lstinline{p$_0$} to \ls{h}. We then
group each regular head with the regular nodes that designate it as
clusterhead: each contains at least $k+1$ regular nodes, {\em i.e},
$rn \ge (k+1)rh$.

Now, we have two cases. If the root is tall, with %
\lstinline{($\alpha$ (g $\r$) $\ge$ $k$)}, every node in $\clrdom$ is
a regular head, \textit{i.e.} is in \ls{RegHeads} and every node is
regular, in \ls{RegNodes}. Otherwise, the root is short and every
clusterhead is a regular head except the root and at least one node is
not regular, namely the root. These two cases yield the following
lemma:
\begin{lstlisting}
  Lemma split_counting_cases: 
    $|\clrdom|$ = $rh$ $\wedge$ $n$ = $rn$ $\vee$ $|\clrdom|$ = $1 + rh$ $\wedge$ $n$ $\geq$ $1 + rn$.
\end{lstlisting}
The proof of the lemma first uses the results on cardinalities, in
particular disjoint union between the singleton containing the root
and the set of regular nodes (resp. regular heads)) and then the above
case analysis.
The main theorem that proves \ls{count_OK} is then just a case
analysis from this lemma.


\section{Conclusion}
\label{sec:ccl}

We proposed a general framework to build certified proofs of
self-stabilizing algorithms. To achieve our goals, we developed
general tools about potential functions, which are commonly used in
termination proofs of self-stabilizing algorithms. We also proposed a
library dealing with cardinality of sets.  We apply our framework to
prove that an existing algorithm is silent self-stabilizing for its
specification and we show a quantitative property on the output of
this case study.

In future work, we expect to certify more complex self-stabilizing
algorithms. Such algorithms are usually designed by composing more
basic blocks. In this line of thought, we envision to certify general
theorems related to classic composition techniques such as collateral
or fair compositions.

Finally, we expect to use our experience on quantitative properties to
tackle the certification of time complexity of stabilizing algorithms,
{\em aka.} the stabilization time.


\bibliographystyle{plain}
\bibliography{biblio}

\end{document}